\documentclass[a4paper,fleqn,usenatbib,useAMS]{mnras}
 \usepackage{graphicx}
 \usepackage{hyperref}
\usepackage{amssymb}

\title[Diagnosing pulsar winds in millisecond pulsar systems]{Diagnosing pulsar winds in black-widow, redback, and other binary millisecond pulsar systems}
\author[A. Zilles, K. Kotera, R. Rohrmann, L. Althaus]
{Anne Zilles$^{1}$\thanks{E-mail:anne.zilles@iap.fr }, 
Kumiko Kotera$^{1}$\thanks{E-mail:kotera@iap.fr }, 
Rene Rohrmann$^{2}$
and Leandro Althaus$^{3}$\\
$^{1}$Institut d'Astrophysique de Paris, CNRS-Sorbonne Universit\'e, 98 bis boulevard Arago, F-75014 Paris, France\\
$^{2}$Instituto de Ciencias Astron\'omicas, de la Tierra y del Espacio (CONICET-UNSJ), Av. Espa\~na Sur 1512, 5400 San Juan, Argentina\\
$^{3}$Facultad de Ciencias Astron\'omicas y Geof\'isicas, IALP-CONICET, Universidad Nacional de La Plata, Paseo de Bosque S/N,\\
{  }1900 Plata, Argentina}

\begin{document}

\date{Accepted XXX. Received YYY; in original form ZZZ}

\pubyear{2019}

\pagerange{\pageref{firstpage}--\pageref{lastpage}} \pubyear{2019}

\maketitle

\label{firstpage}

\begin{abstract}
Binary systems composed of a recycled millisecond pulsar and a 
stellar companion in close orbit could be excellent sites to diagnose pulsar winds. In such systems, the pulsar outflow irradiates and heats up the companion atmosphere, which can lead to the observation of strong day/night modulations in temperature. We demonstrate with particle shower simulations that the particle energy of the wind affects the heating depth in the atmosphere: the wind heat can be deposited above or below the photosphere, leading to different signatures in the observed spectra. We apply our method to four specific systems: We find that systems with cool night side companions showing strong temperature variations can give interesting lower limits on the particle energy in the winds. In particular, if the companion night side of PSR B1957+20 were to be suddenly irradiated, deep heating would only take place if particles with energy $>100$\,TeV were present. Observational evidence of deep heating in this system thus suggests that i) such particles exist in the pulsar wind and/or ii) binary evolution non-trivially takes the companion to the observed temperature asymmetry.
Besides, the observed temperature difference can be maintained only with particle energies of the order of $100\,$MeV. 
\end{abstract}

\begin{keywords}
astroparticle physics, pulsars: general,  acceleration of particles
\end{keywords}


\section{Introduction}

Pulsars lose their energy via electromagnetic cooling principally, under the form of an outflow. The existence of this wind is revealed by its interactions with the interstellar medium and the supernovae debris, observed as pulsar wind nebulae. The wind should be dominantly composed of Poynting flux close to the star, and of relativistic particles at the nebula (e.g., \citealp{Kirk09} for a review on this so-called ``$\sigma-$problem''). 

But the dissipation from electromagnetic to kinetic energy is uncertain, and more generally, there are ongoing debates about the energy, the nature, the structure, formations and evolution of this outflow.

Binary systems where a recycled millisecond pulsar wind impinges on the atmosphere of its companion, could turn out to be a unique laboratory to diagnose the nature and energy of the outflow. More than 60 such systems, including the so-called {\it black widows} and {\it redbacks}, were discovered over the last decade, thanks to follow-ups of Fermi sources \citep{Li18, Strader18, ApatrunoURL}. 
Black widows are millisecond pulsar binaries with low-mass ($M_{\rm c}\sim 0.05\,M_\odot$) evaporating  companions. They are typically distinguished from redback systems that have heavier companions ($M_{\rm c} \gtrsim 0.2\,M_\odot$) \citep{Roberts2013}.
These systems being compact, with orbital periods of $<1$ day, they enable to scan the wind at different distances, at closer ranges compared to what can be classically explored with the nebula. 

As the pulsar wind impinges a substantial irradiative flux on its companion, it is theoretically expected that the companion be heated and show a strong day/night variation. \cite{Phinney88} made this prediction shortly after the discovery of the original black-widow pulsar B1957+20, and the expected orbital modulation of the thermal emission from the companion was quickly observed by \cite{Fruchter90}. A majority of the observations shows that a non-negligible fraction of the expected pulsar wind flux impinges the companion and gets reradiated (see e.g., \citealp{Stappers96, Stappers99,Reynolds07, vanKerkwijk11,  Breton:2013ess, Li18, Strader18}). 
On the other hand, the companion to some pulsars, such as PSRs J2129-0429 \citep{Bellm:2015dfa}, J1723-2837 \citep{osti_22270898, vanStaden:2016ubf}, J1816+4510 \citep{Kaplan:2013hii}, and J2129-0428 \citep{2013AAS...22115410B}, and for example the non-spider system PSR J0751+1807 \citep{Bassa06}, having a white dwarf companion, present low irradiation temperatures with little or no modulation.

Setting out on these observational premises, we explore the effects of relativistic pulsar winds on their companion atmospheres, as a means to probe their composition. Companion heating by winds dominated by Poynting flux was carefully examined in \cite{Kotera2016}. Hence, here we assume that the outflow is mainly loaded in high-energy particles and photons. Assuming that this particle flux impinges vertically on the atmosphere, we estimate the depth of energy deposition in the companion atmosphere. At first order, the efficient heating of the companion should depend on whether the wind energy is deposited above or below the photosphere. Energy deposition below the photosphere offers the possibility to heat the inner atmosphere and increase its observed temperature. Above the photosphere, energy deposition can lead to shallow heating, that could be probed by changes in the stellar spectrum and emission/absorption lines. 

If the companion is tidally locked to the pulsar, as one naturally expects in these systems, the deposited energy should illuminate only one side of the star.  If the companion is not tidally locked, a comparison of the radiative time at the heating depth with the rotation period helps to assess whether the deposited energy will illuminate only one side of the star, or can be distributed over the entire surface.

We calculate the heating depth with the numerical particle-shower simulation tool \textsc{Geant4}~\citep{GEANT}, taking into account the atmosphere column densities and all microscopic processes related to particle interactions and cooling. We compare our findings with several observational examples to draw  conclusions on the parameters of the primary particles composing the pulsar winds. 

In Section~\ref{section:basics}, we recall basic observational and theoretical elements related to millisecond pulsar companion irradiation. We present the \textsc{Geant4} simulation set-up and the numerical results on atmospheric heating depth in Section~\ref{section:showers}. We apply our outputs to observed binary systems in Section~\ref{section:application} and draw our interpretations and conclusions on the composition of the wind in Section~\ref{section:conclusion}.

\section{Companion irradiation: observations and theoretical background}\label{section:basics}

About 300 pulsars among roughly 2600 pulsars listed in the ATNF Pulsar Catalogue \citep{Manchester05} are identified as millisecond pulsars (MSPs). Most of the observed MSPs are found in binary systems, where they are believed to have been spun up by transfer of mass and angular momentum from the companion. So far, the observed numbers indicate that $10-30\%$ of the MSPs are black-widow or redback systems, with low mass companions (D. Smith, private communication). Black widow pulsars have companions of mass $0.01-0.05$\,$M_\odot$ and orbital period less than $P_{\rm{b}} = $10$\,$h \citep{vanKerkwijk:2004tm}.
 Redback pulsars have binary companions with higher masses ($\gtrsim 0.2\,M_\odot$) and orbital periods of less than a day ($P_{\rm{b}} < $1$\,$d) \citep{Roberts2013}.

\subsection{Observations of companion temperature modulations}\label{observations}

Optical observations of the companion can help determine the parameters of the companion (e.g., \citealp{vanKerkwijk:2004tm, Romani:2011xf}). The radial velocity curve and the atmospheric parameters of bright companions can be obtained through optical spectroscopy, to determine the mass of the companion and the pulsar. Combined with phase-resolved spectroscopy, this can be used to determine the component masses (see \citealp{Breton:2013ess, Linares:2018ppq}).  
The modelling of the orbitally modulated light curves can then constrain the companion temperature variation, the orbital inclination and the irradiation efficiency of the companion by the pulsar wind (see \citealp[references therein]{Breton:2013ess}).

\begin{figure}
\centering
\includegraphics[width=0.48\textwidth]{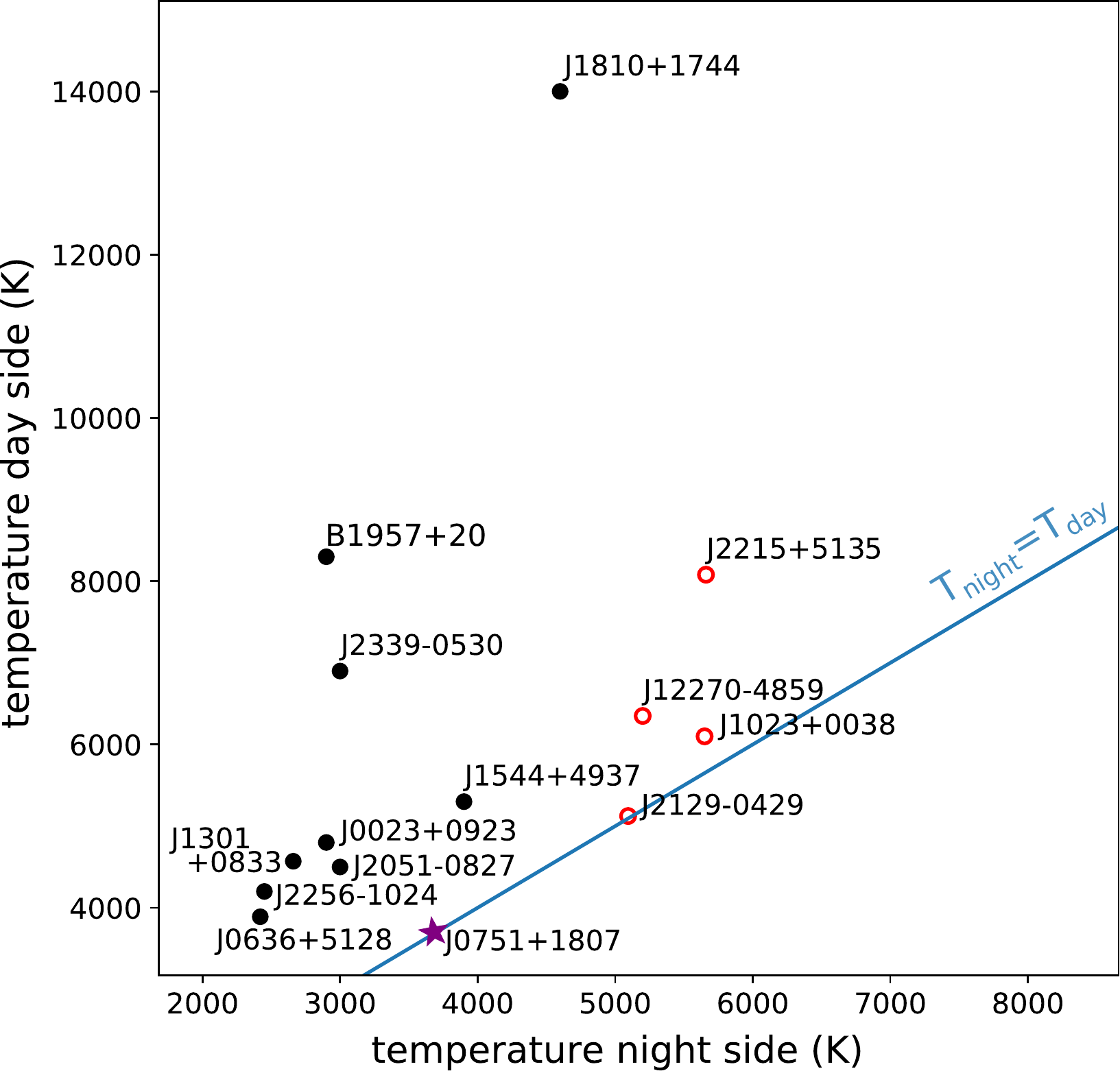}
\caption{Observed temperatures for day and night sides for several BW and RB systems. The black dots represents black-widow systems, the red red-back systems, the system J0751+1807, marked by a star, has a white-dwarf companion; detailed informations are given in Tab.~\ref{tab:pulsars}. The blue line represents equivalent temperatures for the two sides.}
\label{fig:scales}
\end{figure}

In most of these systems, the pulsar wind impinges a substantial irradiative flux on its companion. It is thus  expected that the companion be heated and show a strong day/night variation. Although such a modulation has been observed in many systems (see Fig.~\ref{fig:scales}), detailed information on the companion temperature of the day and the night sides have only been published for a few systems. A non-exhaustive list of systems for which temperatures measurements are available can be found in Table~\ref{tab:pulsars}, and their temperatures are represented in Figure~\ref{fig:scales}.

Companion temperature measurements are difficult, as the optical light curves from which they are inferred can be affected by various effects, such as tidal distortion, migrating star spots, etc., and require a precise derivation of the effective temperature of the star, via spectroscopy (e.g., \citealp{Strader18,Cho2018}). The temperature modulation can also be interpreted as a probe of the presence of intrabinary shocks (e.g., \citealp{Cho2018}). 
 Some benchmark examples are treated as applications of our results in Section~\ref{section:application}.

\begin{table*}
\caption{Compilation of black-widow (first block of lines) and redback (second block of lines) systems and one millisecond pulsar-white dwarf system, which have measured companion temperatures. We list the pulsar period $P$, energy loss rate $\dot{E}$, the orbital period $_{\rm b}$, the companion's mass $M_{\rm C}$ and the observed temperatures of the day $T_{\rm day}$ and the night side $T_{\rm night}$, respectively, the irradiation temperature $T_{\rm irr}= (T_{\rm day} ^4 - T_{\rm night}^4)^{1/4}$~\citep{Bellm:2015dfa} and the orbital separation $a$ between pulsar and companion.
Uncertainties from single measurements are not listed below. References are not exhaustive: we mainly quote the seminal observation paper and the reference giving the companion temperature modulation.}
\label{tab:pulsars}
\begin{center}
\begin{tabular}{l c c c c c c c c  p{5.5cm}}
\hline
Pulsar               & $P$    & $\dot{E}/10^{34}$  & $P_{\rm b}$ & $M_{\rm C}$     & $T_{\rm day}$ & $T_{\rm night}$ & $T_{\rm irr}$  & $a$ & References, e.g. \\
                     & [ms] & [erg s$^{-1}$]      & [h]   & $M_\odot$ & [K]       & [K]  & [K] & $R_\odot$ &   \\     
\hline 
B1957+2048             & 1.61 & 11 & 9.2 & 0.021 & 8300 & 2900 &  8269 & 2.5 &  \cite{van_Kerkwijk_2011,Huang_2012, Khechinashvili:2000as}\\
J2051-0827           & 4.51 & 0.5 &2.4 & 0.027 & 4500 & $<$3000 &  $>$4259 & 1.0 & \cite{Khechinashvili:2000as,Lyne2013, Khechinashvili:2000as} \\
J0023+0923           & 3.1 & 1.51 & 3.33 & 0.017 & 4800 & 2900 &  4631 & 1.27 &  \cite{gentile_2012,Breton:2013ess}\\
J2256-1024           & 2.3 & 3.95 & 5.11 & 0.030 & 4200 & 2450 &  4073 & 1.69 &   \cite{gentile_2012,Breton:2013ess} \\
J1301+0833           & 1.84 & 5.0 & 6.5 & 0.024 & 4570 & 2660 &  4433 & 2.2$^*$ &   \cite{2016ApJ...833..138R,Li:2014tka}\\
J1544+4937         & 2.16 & 1.15 & 2.9 & 0.017 & 5400 & 3901 &  4987 & 1.2 &   \cite{Bhattacharyya:2013ora, Tang:2014nja}\\ 
J1810+1744           & 1.7 & 3.97 & 3.56 & 0.07 & $\sim$14000 &  $\sim$4600 & $\sim$13959 & 1.33 &  \cite{gentile_2012,Breton:2013ess} \\ 
J2339-0530			& 2.88 & 2.3 & 4.6 & 0.075 & $\sim$ 6900 & $<$3000 &  $>$6838 & 1.71 &  \cite{Abdo_2009,Romani:2011xf,Pletsch2015} \\
J0636+5128			& 2.87 & 2.3 & 0.07 & 0.019 & 3890 & 2420 &  3735 & 0.8$^*$ &   \cite{Draghis:2018mkh,Kaplan:2018bwp} \\ 
\hline 
J1023+0038           & 1.69 & $\sim$5 & 4.8 & 0.2 & 6100 & 5650 &  4373 & 1.65 &  \cite{Thorstensen_2005,2009Sci...324.1411A, Breton:2013ess}\\
J2215+5135           & 2.6 & 5.29 & 4.14 & 0.33 & 8080 & 5660 &  7542 & 1.53 &   \cite{gentile_2012,Breton:2013ess, Linares:2018ppq}\\
J2129-0429           & 7.61 & 34.6 & 15.2 & 0.44 & 5124 & 5094 &  2000 & 3.9$^*$ &  \cite{Bellm:2015dfa,2013AAS...22115410B}\\ 
J12270-4859			&1.69 & 9.0 & 6.91 & $>$ 0.01 & $\sim 6350$ & $5200$ &  $\sim$5469 & 2.1$^*$ &  \cite{Baglio:2016pow,Sandoval:2017bog}\\
\hline
J0751+1807           & 3.48 & 0.8 & 6.3 & 0.12 & 3700 & 3700 & --- & 1.9$^*$ &  \cite{Bassa06,Fortin:2014ufa}\\ 
\hline
\hline
\end{tabular}
\end{center}
\begin{flushleft}
 \footnotesize{$^*$ approximated using: $a=0.6R_\odot(P_{\rm b}/{\rm h})^{2/3}(M_{\rm PSR} /1.5M_\odot)^{1/3}$, with $M_{\rm PSR}$ as the pulsar mass.}
\end{flushleft}
\end{table*}

\subsection{Nature and energetics of pulsar winds}

Most numerical applications in this section will assume values close to those observed for PSR 1957+20, one of the best studied black widow systems (see Table~\ref{tab:pulsars} and Section~\ref{section:application}).
All numerical quantities are denoted $Q_x\equiv Q/10^x$ in cgs units unless specified otherwise.

\subsubsection{Composition of pulsar winds}

Observations from the Fermi space telescope have revealed that millisecond pulsars, including those in black widows and redbacks, have a GeV gamma-ray luminosity which is a significant fraction $f_\gamma$ of the pulsars spin-down luminosity. The inferred $f_\gamma$ ranges from 0.01 to 5, with 0.1 being a typical value \citep{2014arXiv1407.5583C}.  
The flux per log photon energy $E$ $E^2 {\rm d}N/{\rm d}E\propto E^{2-\alpha}\exp(-E/E_{\rm c})$ with $2-\alpha\sim 0.4$ and $E_{\rm c}\sim 4\,\mbox{GeV}$, so most of the gamma-ray power is emitted in $1-3\,\mbox{GeV}$ gamma-rays \citep{2014arXiv1407.5583C}.  
This gamma-ray flux represents a {\em minimum} source of heating of the pulsar-facing side of the pulsar companion, with a very simple illumination function: the gamma-rays, often observed to be modulated at the pulsar pulse period, are believed to come from in or near the light-cylinder, typically $<10^{-4}$ the distance to the companion and thus are effectively a point source.

The rest of the pulsar spin-down energy is carried by a combination of electrons, positrons and ions accelerated by the large induced voltages (see discussions in \citealp{Arons03,Fang12,Kotera15,Lemoine15} on ion injection and acceleration in pulsar winds), and by Poynting flux (as is considered in \citealp{Kotera2016}). The achievable energies of these particles is estimated in the following.

\subsubsection{Particle acceleration in pulsar winds}

The energy loss rate of a pulsar with moment of inertia $I=10^{45}I_{45}\,$cgs, rotation period $P=10^{-3}P_{-3}\,$s, and period derivative $\dot{P}=10^{-20}\dot{P}_{-20}\,$s\,s$^{-1}$ reads (e.g., \citealp{Shapiro83}):
\begin{equation}
\dot{E}_{\rm p} = I(2\pi)^2\frac{\dot{P}}{P^3}\sim 3.9\times 10^{35}\,{\rm
 erg\, s}^{-1}\,I_{45}\dot{P}_{-20}P_{-3}^{-3}\ .
\end{equation}
In force-free aligned pulsar magnetospheres, it can be calculated that the combination of the strong magnetic moment and fast rotation can induce voltage drops of magnitude: 
\begin{equation}
\Phi = \left( \frac{4\dot{E}_{\rm p}}{c}\right)^{1/2} \sim 2.2\times 10^{15}\,(I_{45}\dot{P}_{-20}P_{-3}^{-3})^{1/2}\,{\rm V}\ .
\end{equation}
Particles of charge $Z$ and mass number $A$ experiencing a fraction $\eta=0.3\,\eta_3 $ (for ion-dominated winds model for the the Crab pulsar, \citealp{Hoshino92}) of these voltage drops can be accelerated to Lorentz factor:
\begin{equation}
\gamma = \eta \frac{Ze}{m_{\rm i}c^2}\Phi \sim 6.9\times 10^5\frac{Z}{A}(I_{45}\dot{P}_{-20}P_3^{-3})^{1/2}\ .
\end{equation}
In principle, other pulsar configurations allow to tap the rotational energy of the pulsar into the wind, for example in the equatorial current sheet, and accelerate particles up to these energies (see e.g., \citealp{Kirk09}).

\subsubsection{Energy flux of pulsar winds intercepted by the companion}
The energy flux in the pulsar wind at distance $r$ large compared to the pulsar light cylinder radius, $R_{\rm L} = cP/(2\pi)\sim 4.8\times 10^8\,$cm\,$P_{-3}$, can then be written \citep{Arons93}:
\begin{equation}\label{eq:FwEr}
F_{\rm w} = \frac{\dot{E}_{\rm p}}{4\pi f_{\rm p}r^2}=\frac{I\pi}{f_{\rm p}r^2}\frac{\dot{P}}{P^3}\ ,
\end{equation}
where we noted $f_{\rm p}=\Delta\Omega_{\rm p}/4\pi$ the fraction of the sky into which the pulsar wind is emitted.

The companion can intercept a fraction $f$ of this flux, provided that it falls in the wind beam. We note the semi-major axis of the companion orbit $a=0.6R_\odot(P_{\rm b}/{\rm h})^{2/3}(M/1.5M_\odot)^{1/3}$, with $P_{\rm b}$ the binary period in hours. The characteristic age of the pulsar is noted $\tau_{\rm c}=P/2\dot{P}$.  
The ratio of the incident flux in the pulsar wind on the ``day" side of the companion to the natural cooling flux on the ``night" side (with temperature $T_{\rm night}$) would thus be
\begin{eqnarray}\label{eq:flux}
f_e &=& \frac{fF_{\rm W}}{\sigma_{\rm T} T_{\rm night}^4} = \frac{f}{f_{\rm p}}\frac{\dot{E}_{\rm p}}{4\pi a^2\sigma_{\rm T} T_{\rm night}^4}\label{eq:fe}\\
&\sim& 5\times 10^5 ff_{\rm p}^{-1}P_{-3}^{-2}\tau_{\rm c,Gyr}^{-1}M_{1.5}^{-2/3} 
\left(\frac{P_{\rm b}}{1 {\rm h}}\right)^{-4/3}\left(\frac{T_{\rm night}}{10^3\,{\rm K}}\right)^{-4}\nonumber \ ,
\end{eqnarray}
where we have assumed an isotropic wind emission, and a full interception fraction for the numerical estimate. This calculation shows that, at first order, less than $ff_{\rm p}^{-1}\sim0.01\%$ of the pulsar wind flux is required to provide the energetics to double the temperature of the day side. The wind thus provides largely enough energy to heat up the companion atmosphere. 
 
The irradiation efficiency can also be measured by the ratio between the difference in radiated flux between the day and night sides and the pulsar wind flux on the day side:
\begin{eqnarray}\label{eq:radeffiency}
\eta_{\rm irr} &=& \frac{f_{\rm p}}{f}\frac{4\pi a^2\sigma_{\rm T} T_{\rm irr}^4}{\dot{E}_{\rm p}}\label{eq:etairr}\\
&\sim& 1\% f^{-1}f_{\rm p}P_{-3}^{2}\tau_{\rm c,Gyr}^{1}M_{1.5}^{2/3} 
\left(\frac{P_{\rm b}}{1 {\rm h}}\right)^{4/3}\left(\frac{T_{\rm irr}}{8000\,{\rm K}}\right)^{4}\nonumber \ ,
\end{eqnarray}
where the irradiation temperature, defined as 
\begin{equation}
T_{\rm irr} = (T_{\rm day}^4-T_{\rm night}^4)^{1/4}
\end{equation}
provides a more sensitive estimate to the actual temperature modulation. $\eta_{\rm irr}$ gives an estimate of how much energy from the pulsar wind was channelled into heating the companion atmosphere.

For this study we assume that a sufficient fraction of the pulsar rotational energy impinges the companion atmosphere under the form of high energy photons or particles, without experiencing drastic energy losses. The details of how the particles propagate and interact between their acceleration site and the companion atmosphere are not considered. The energy reached at the companion atmosphere is left essentially as a free parameter. Deflection in the companion magnetic field could affect the propagation of the lowest energy particles in the atmosphere: this point is discussed in Section~\ref{section:magnetosphere}.

\subsection{Day/night effects and atmosphere heating}
From Eq.~(\ref{eq:fe}), we expect the companions of millisecond pulsars with orbital periods less than about 10 hours to be efficiently heated by the incident pulsar wind. As listed in Tab.~\ref{tab:pulsars}, most of the given examples show indeed a clear orbital modulation. By contrast, the companion to PSR J0751+1807, which should have a day-side twice as bright as the night side ($f_e\sim 1$) given by the irradiative flux of the pulsar wind, presents no detectable modulation.

The ratio of the heating column depth to the photospheric column depth: 
\begin{equation}\label{form:ratio}
\xi=\frac{\Sigma_{\rm heat}}{\Sigma_{\rm phot}}
\end{equation}
is a useful diagnostic of the expected behavior of companions subject to irradiation.  Deep heating ($\xi \gg 1$), which is seen to occur in hot atmospheres ($>4000$K), will produce the usual photospheric temperature profile decreasing outwards, leading to the usual absorption lines, but at the new (irradiated) temperature.

Shallow heating ($\xi < 1$), which is seen to occur in cool atmospheres, should produce
a flat or inverted temperature profile at or above the photosphere, leading to
weakening of the lines, or even the appearance of the lines in emission.

The second useful number is the radiative time at the heating depth, defined as the ratio of the heat flow by diffusion and the radiated energy,
\begin{equation}
  t_{\rm rad}=\frac{c_p T\Sigma_{\rm heat}}{\sigma T^4},
\label{eq:radtime}
\end{equation}
where $c_p\simeq 2.5 k/\mu$, with
  $k$ the Boltzmann constant and $\mu$ the mean mass per particle ($\sim m_H$ for neutral hydrogen, and $m_H/2$ for ionized hydrogen). $\sigma T^4$, with $\sigma$ as the Stefan-Boltzmann constant, corresponds to the total flux radiated through the layers above the heating layer, with $T$ the temperature of the atmosphere. 
As calculated in Section~\ref{section:application}, the values of $t_{\rm rad}$ range from days for incident TeV particles, to less than minutes for $< \mbox{GeV}$ ones.

\section{Depth of the energy deposition in companion atmospheres}\label{section:showers}

To understand the dependency of the energy deposition on the primary particle type, we estimate in this section the heating depth of the companion atmospheres, as a function of the nature of the primary and its initial energy. The latter parameters determine the development of the induced particle shower and therefore the depth of the energy deposition. 
To express the position of the maximum heating in the companion's atmosphere, in other words where most of the energy is deposited, independent of a specific density profile $\rho(z)$, one can determine the transversed column density 
\begin{equation}\label{columndensity}
	 \sum(<z)\equiv \int_0^z \rho(z') dz'.
\end{equation}
The column density at which the energy deposition of the particle shower reaches its maximum $\Sigma_\mathrm{max}(<z)$ will be called the heating depth in the following and corresponds to $\Sigma_{\rm heat}$ in Eq.~(\ref{form:ratio}).

\subsection{Particle air-showers}\label{sec:particleshowers}
When a high-energetic particle enters a companion atmosphere, it will induce a cascade of secondary particles ~\citep{CRBook}.
High-energetic electrons, positrons and photons initiate electromagnetic showers, containing millions of charged particles of lower energies. At high energies (above a few MeV, below which photoelectric effect and Compton scattering are dominant), photons interact with matter primarily via pair production, convert into an electron-positron pair, while interacting with an atomic nucleus or electron in order to conserve momentum. 

While for heavy particles (proton, muon, pion) of an energy $E_0$ energy losses happen mainly via collision/scattering with atoms leading to an excitation, the energy loss for electrons and positrons is dominated by the emission of photons, called bremsstrahlung (for photons pair production, respectively), for particle energies above $E> 370\,$MeV. Below this so-called critical energy $E_c$ the shower development is dominated by ionisation and scattering, rather than by production of further particles \citep{PhysRevD.98.030001}. The particles lose their energy and the shower ``dies out" by absorption of the particles in the atmosphere. The depth of the maximum energy deposition can be approximated by 
\begin{equation}\label{heat_em}
\Sigma_\mathrm{max} (<z) \sim X_0 \ln(E_0/E_c)
\end{equation}
following the Heitler toymodel~\citep{CRBook} which can be used to estimate when the showers reaches its maximum particle number. The parameter $X_0=63\,$g/cm$^2$ represents the radiation length in the Hydrogen-dominated atmosphere.

Hadrons as primary particles of an energy $E_0$ interact with a nucleus of the atmosphere via a nuclear reaction after propagating through a mean column density $X_h= 35\,$g/cm$^2$ (nuclear interaction length) for a Hydrogen-dominated atmosphere and produce secondary hadrons. Each of them will continue interacting or will decay. Here, most of the produced particles in hadronic interactions are pions and kaons which can decay into muons and neutrinos before interacting. Muons are more penetrating with radiation length $X_\mu \sim 500\,$g/cm$^2$, and decay after travelling $0.66\gamma\,$km, typically more than the companion’s atmospheric scale height, while neutrinos can partially escape. Neutral pions, on average $1/3$ of the produced pions, will dissipate their energy in the form of electromagnetic showers, which will dominate the shower development and therefore the energy deposition, by decaying into photons~\citep{shower_rpp}. Gradually, mainly electromagnetic particles are produced. Within the Heitler toymodel one assumes that the total number of particles $N$ increases until it fulfils $E_0/N < E_c$. Below this critical energy $E_c \sim 100\,$GeV~\citep{CRBook}, which depends on altitude and density, the particle decays rather than interacts. Finally, ionization losses degrade the energy into heat, low-energetic particles get absorbed and the shower ``dies out". This column depth at which the maximum number of particles is reached and the heat is mostly deposited is thus approximately
\begin{equation}\label{heat_hadr}
\Sigma_{\rm max}(<z) \sim X_h \ln(E_0/E_c) \ .
\end{equation}
 
From the given scaling and the values for $X_h$ and $X_0$, it appears that electromagnetic showers of the higher energies will penetrate deeper into the Hydrogen-dominated atmosphere. Electromagnetic showers also appear to lead to deeper maximum heating, but this simplified model does not account for the production of the electromagnetic subshowers in hadron-induced showers. These electromagnetic showers have dominant effects in the heating depth, as we will see in the numerical simulations. In general, the Heitler model is a  simplistic description of the development of particle showers and can just describe general features. It neglects any interaction with the nuclei in the atmosphere and assumes a constant multiplicity per particle generation.

\subsection{Numerical setup}\label{sec:setup}

We used the toolkit \textsc{Geant4} 10.4 \citep{GEANT} to simulate the particle shower induced in the companions atmosphere and to determine the deposited energy. This toolkit is object-oriented and programmed in C++. It simulates the passage of particles through dense material and treats their propagation and interactions. 
Here, the shower development is simulated by splitting up continuous trajectories of particles into sub-tracks, called steps. All relevant interactions of shower particles are taken into account. 

The use of \textsc{Geant4} requires 3 mandatory user classes: one to specify a particle gun (beam of injected particles with initial conditions), a second defining a detector (physically defined volume in which the interactions will take place), modelling the atmosphere as well as one class to specify a ``Physics List'', namely a list of physical processes that shall be accounted for during the propagation of particles in the detector. 

\begin{description}
\item[\bf Particle gun:] $N_{\rm i}$ particles of a given type are injected with an initial energy $E_{\rm i}$, all in the same direction along the $z$-axis, and at the same position at the edge of the box ($z=0$).

\item[\bf Detector:] The volume in which particles are propagated is a rectangular box of dimensions $(X,Y,Z)$ (with $Z=5\,$km being the depth of the atmosphere and $X,Y$ large enough to fully contain the lateral shower development), filled with hydrogen gas as default setting. For a uniform atmosphere, the gas density is set as constant in that box. To simulate a density gradient, $N_{\rm layers}$ thin volumes as atmosphere layers of size $(X,Y,z)$ are placed in the mother volume, with a constant $z$. The density of gas in each layer is calculated following $\rho(z) = \rho_0\exp(z/h)$, where $\rho_0$ and $h$ depend on the atmospheric structure that we consider. 
Particles are propagated in the detector, and the energy deposited in the target volume is recorded at each step (i.e., at each interaction), for primary and secondary particles.

\item[\bf Physics List:] In this study, we use pre-defined reference physics lists\footnote{\url{http://geant4-userdoc.web.cern.ch/geant4-userdoc/UsersGuides/PhysicsListGuide/fo/PhysicsListGuide.pdf}} containing all electromagnetic and hadronic processes that we need in our framework (multiple scattering, ionisation, bremsstrahlung, Compton scattering, gamma conversion, photoelectric effect, pair production, annihilation). 
Hadronic, photo-hadronic, and lepto-hadronic cascades can be treated with \textsc{Geant4}  up to particle energy $E=100\,$TeV. Note however that only proton-proton interactions are implemented up to $100\,$TeV: hadronic interactions involving heavier ions can only be treated up to 10\,TeV.
\end{description}

To reduce the impact of shower-to-shower fluctuations, we simulated several showers for each energy and type of the primary particle. Furthermore, we adapted the number of simulated primaries according to the particle’s energy. This allows to achieve enough statistics for all energies while keeping the used CPU time reasonable (chosen parameters are given in Table~\ref{tab:table1}). In the simulation of the shower and the calculation of the energy deposition, we take all types of secondary particles and their interactions into account. Furthermore, all effects occurring during the shower development, as for example the ionisation of the surrounding material and a possible feedback on the development, are treated within the simulation.  The results on the heating depth of the following study are based on the simulated distributions of the energy deposition, normalised to the number of primaries and their energy.

\begin{table}
\caption{\textsc{Geant4}  simulation parameters. Number of injected primaries $N_{\rm i}$ at each primary energy $E_{\rm i}$.}
\begin{center}
		\begin{tabular}{ c | c | c | c | c | c | c  }
		  $E_{\rm i}$ (GeV) & $1\,$ & $10\,$ & $100\,$ & $10^3\,$ & $10^4\,$ & $10^5\,$ \\ \hline
      $N_{\rm i}$ & $5\times 10^3$ & $10^3$ & $500$ & $100$ & $20$ & $5$ \\			
		\end{tabular}	\label{tab:table1}
		\end{center}
\end{table}

\subsection{Energy deposition by particle showers in companion atmospheres}

To study the impact of a pulsar wind on its companion, we simulate a particle shower in a hydrogen (H) target with a constant density of $1\,$g/cm$^3$, induced by a high-energetic particle. We read out the deposited energy for each length bin and calculate the transversed column density as given by Eq.~\ref{columndensity}.

\begin{figure*}
    \includegraphics[width=0.45\linewidth]{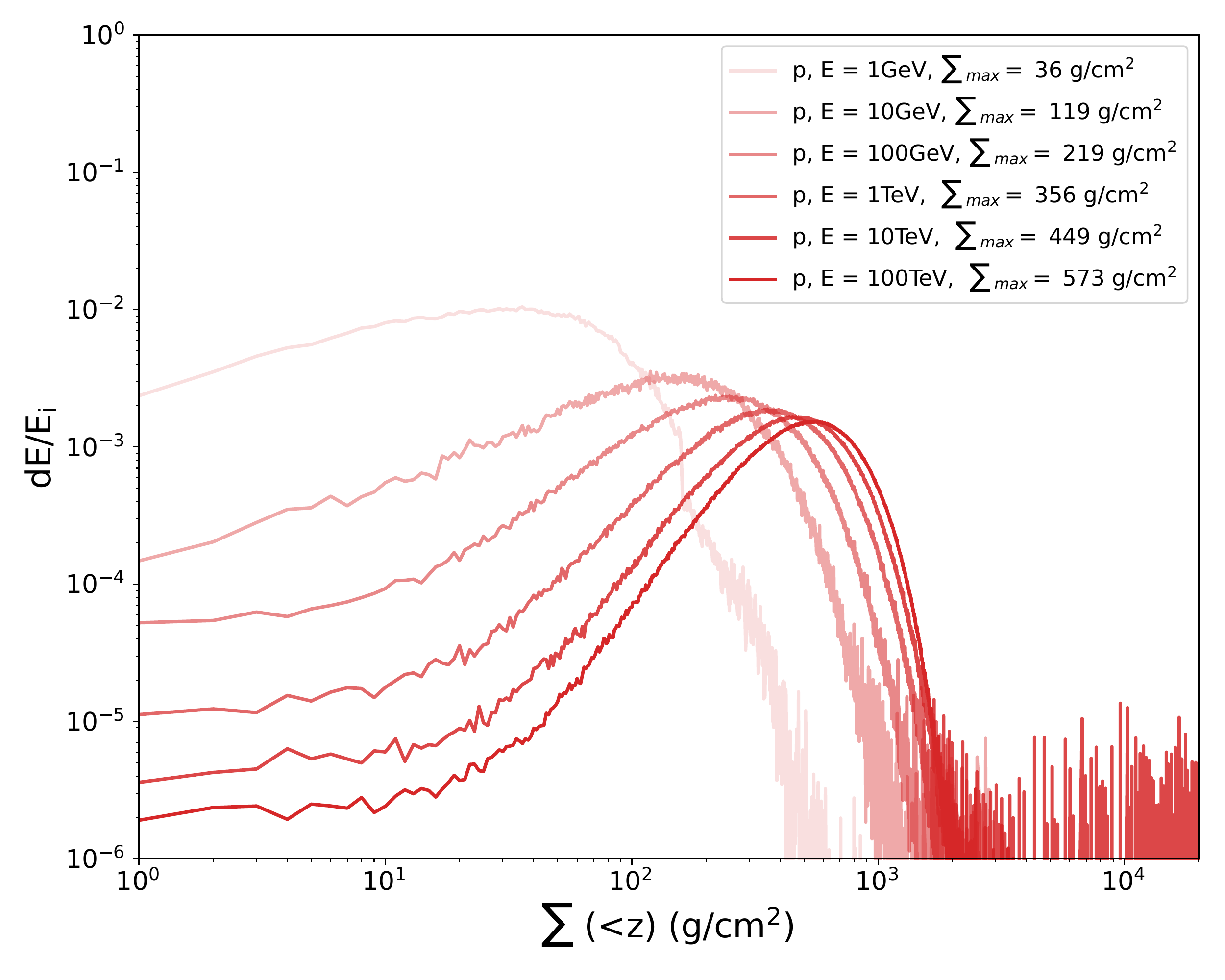}
    \includegraphics[width=0.45\linewidth]{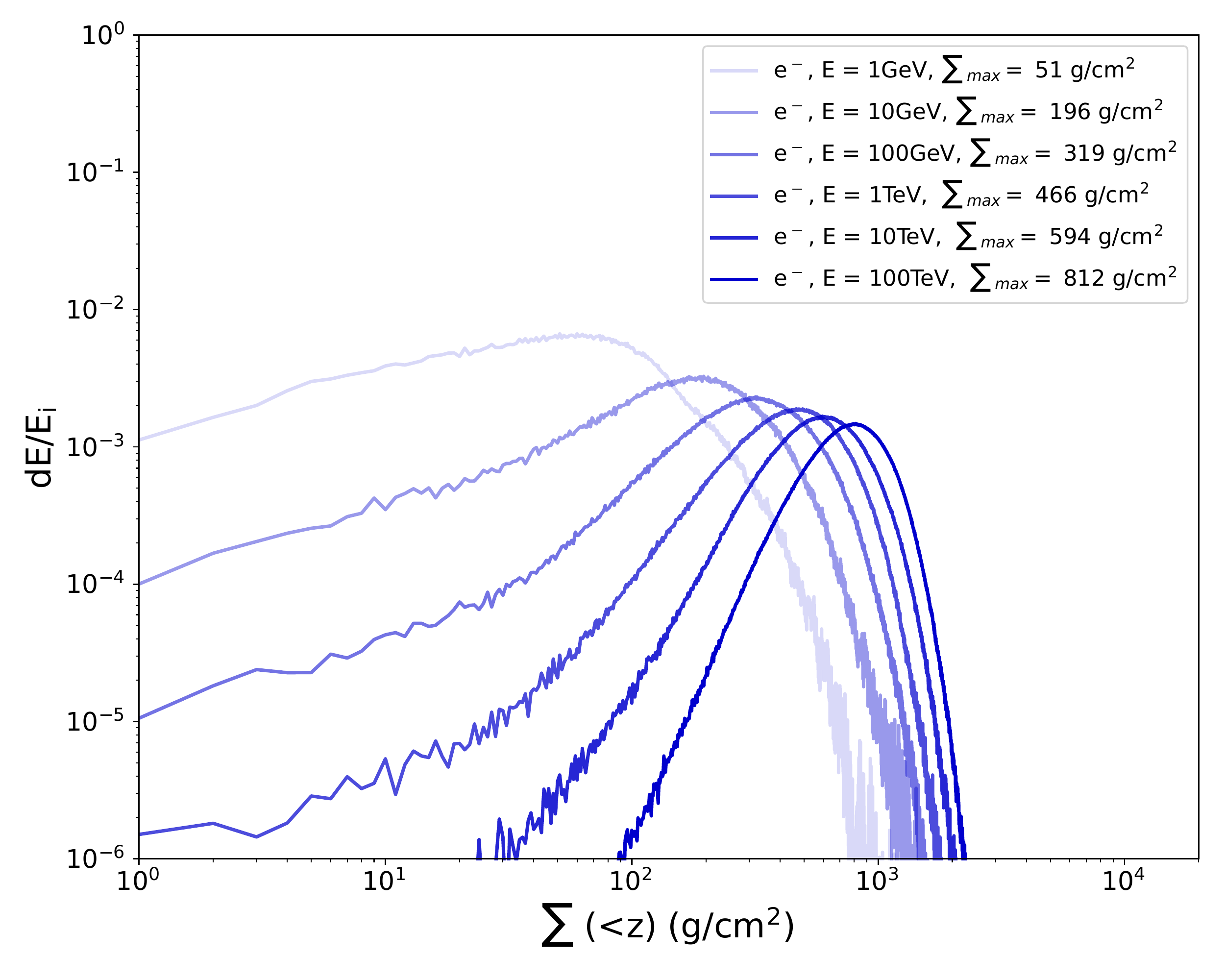}
    \includegraphics[width=0.45\linewidth]{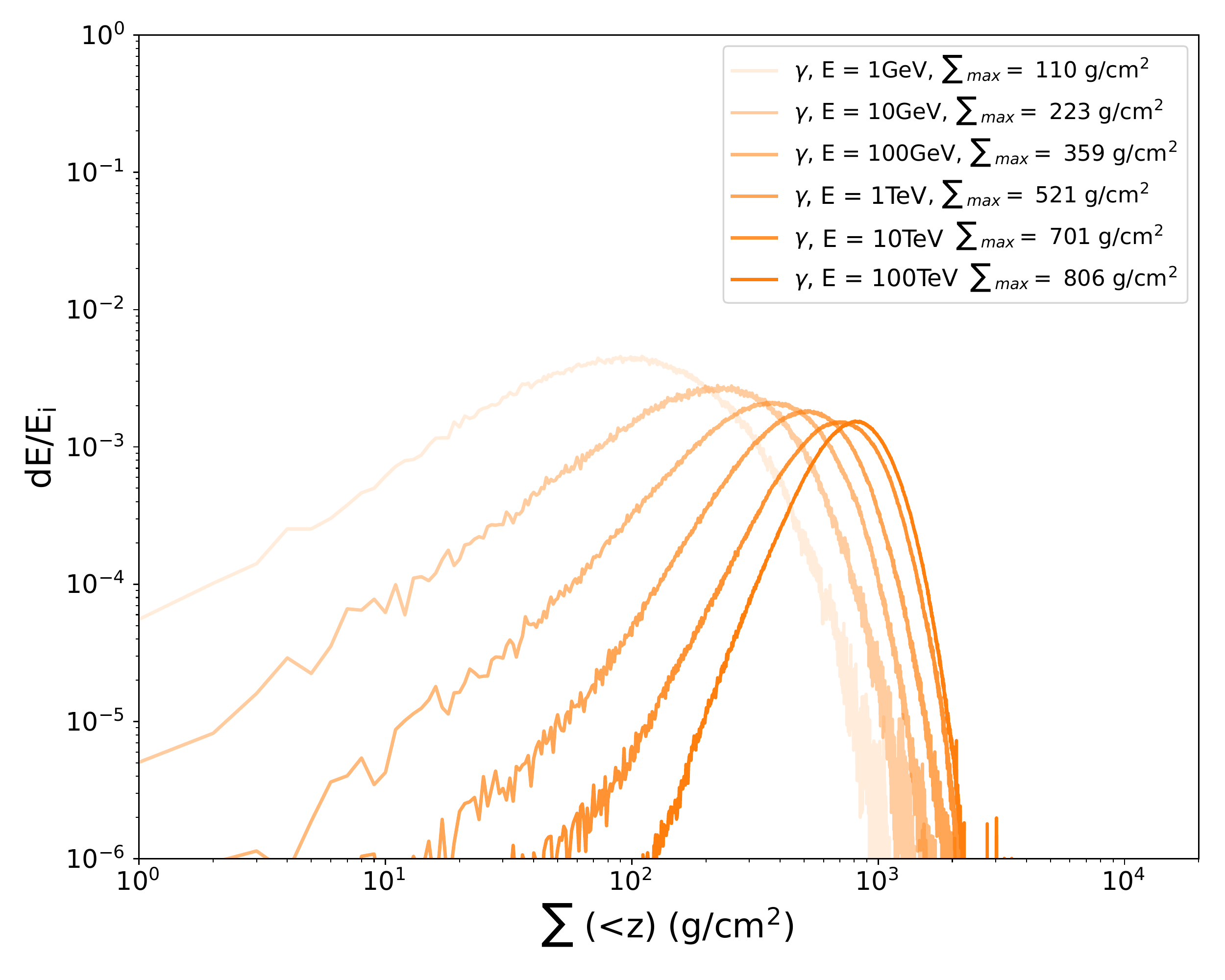}
    \includegraphics[width=0.45\linewidth]{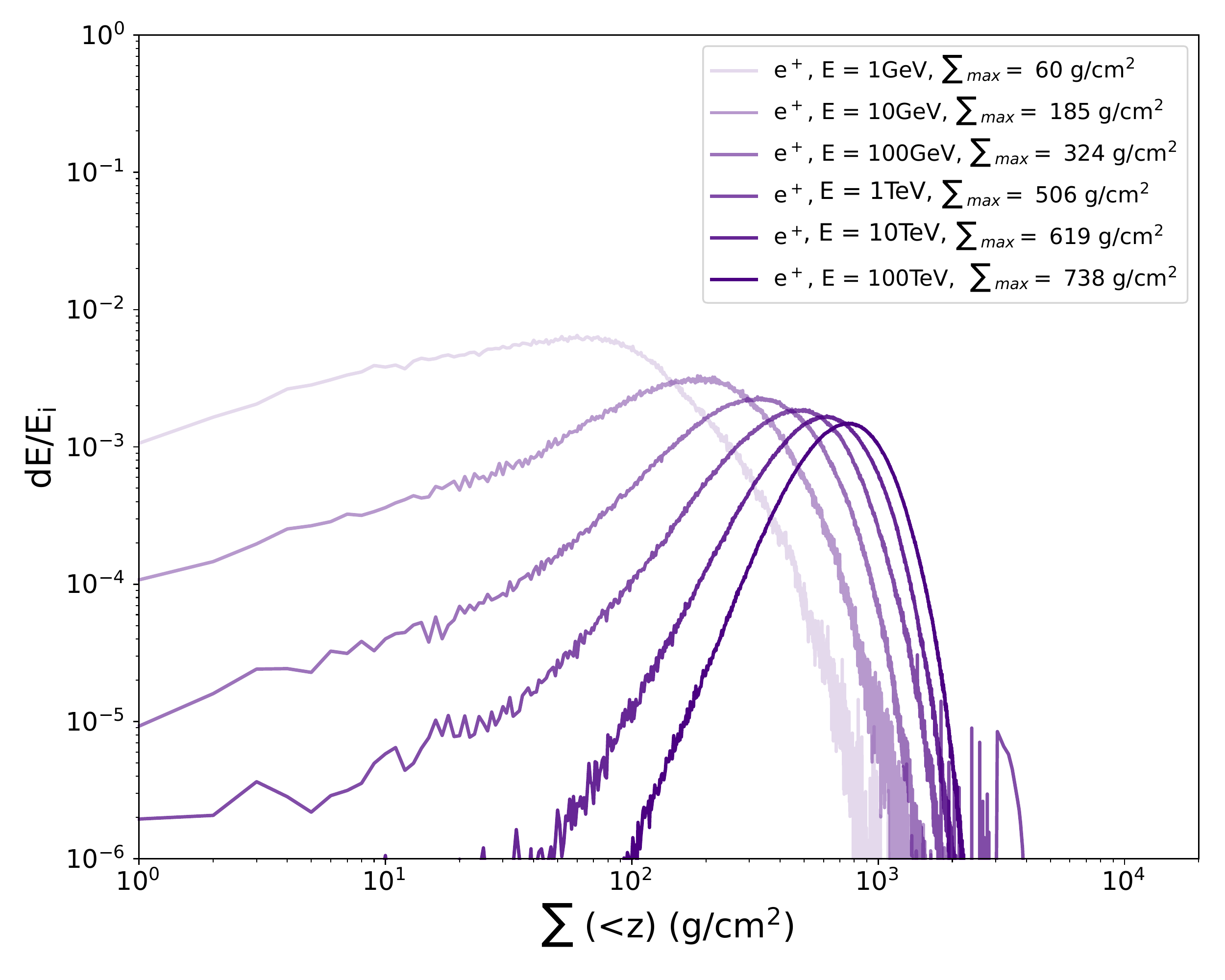}
  \caption{Fraction of the initial energy deposited in the atmosphere, as a function of atmospheric depth, for various primary energies: the position of the maximum energy deposition, also defined as heating depth $\Sigma_\mathrm{max}(<z)$ are given in the label for each primary and energy combination. }
  \label{fig:edep_primaries_energies}
\end{figure*}

The results for the energy deposition for the different types of primaries, gamma ray ($\gamma$), positron ($e^+$), electron ($e^-$) and proton ($p$), and for various primary energies are displayed in Fig.~\ref{fig:edep_primaries_energies}.
One can observe that the maximal energy deposition follows the analytical values found in Sec.~\ref{section:showers} for a Hydrogen atmosphere. Here, the shower development and therefore the energy deposition is dominated by the electromagnetic component. As expected from the scaling of the heating depth with logarithm of the primary energy (compare to Eq. \ref{heat_em} and \ref{heat_hadr}), a higher initial particle energy leads to deeper heating for all primaries since with increasing primary energy more generations of secondaries can be produced in the shower before finally ending.

\begin{figure}
\begin{center}
\includegraphics[width=0.48\textwidth]{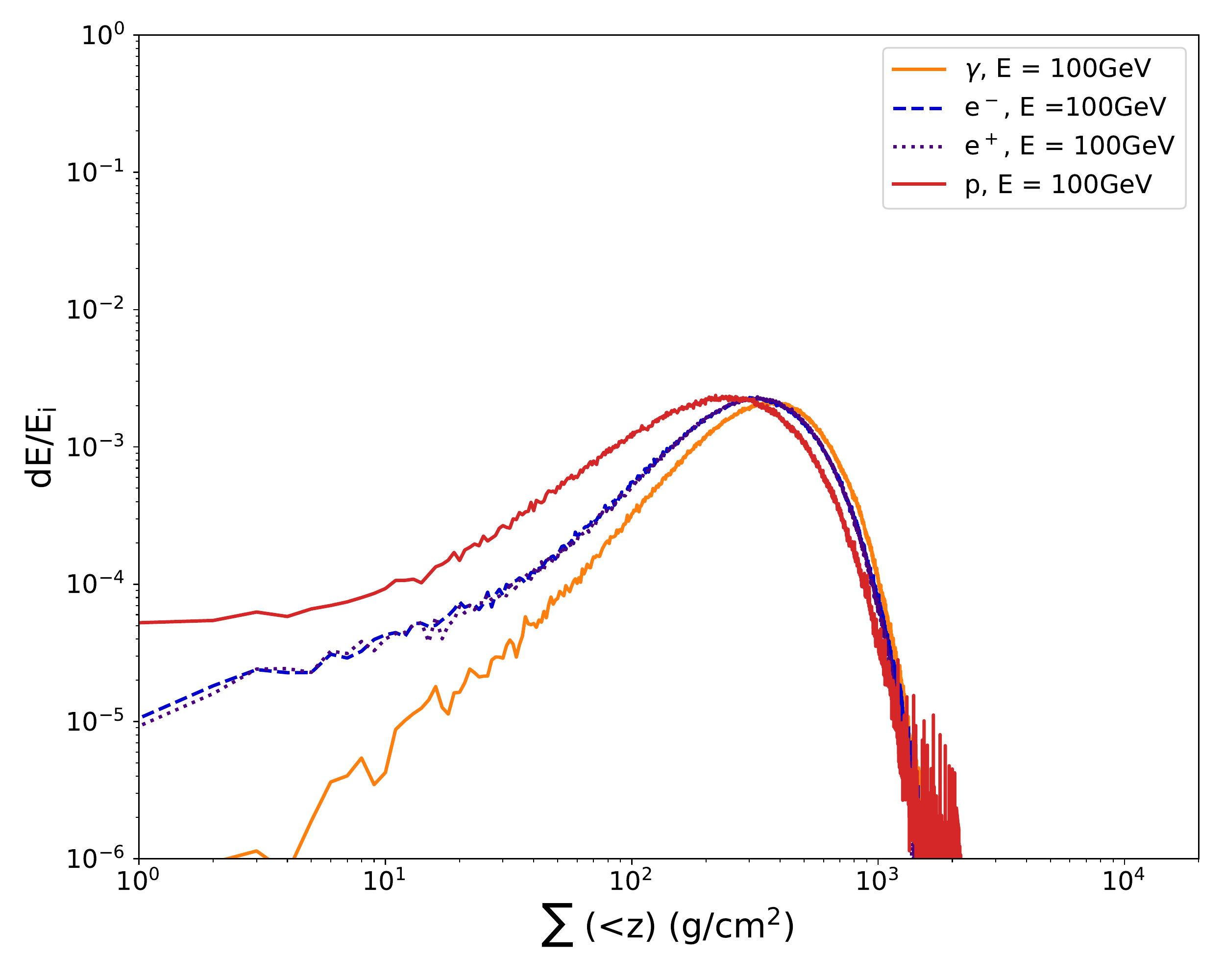} 
\caption{Fraction of the initial energy deposited in the atmosphere, as a function of atmospheric depth, for various primary type with the initial energy of $100\,$GeV.}\label{fig:edep_primaries}
\end{center}
\end{figure}

For a closer look on the heating depth for the various primaries, Fig.~\ref{fig:edep_primaries} shows the fraction of the initial energy deposited in the atmosphere, as a function of atmospheric depth, for an initial energy of $100\,$GeV for the four primaries.

It can be clearly seen that a particle shower induced by a hadron deposits the energy shallower than the showers induced by the leptons or gamma rays. This finally leads to a less deep penetration into the atmosphere and ends in a shallower heating. But the difference in the heating depth for the different primaries with a fixed primary energy is less pronounced so that it does not seem possible to resolve specific primaries given the current resolution of optical observations.
The results for electrons and positrons are consistent since they undergo the same processes. For gamma rays as primaries, the heating will take place slightly deeper in the atmosphere. The leptons mainly responsible for the energy loss have first to be produced in the dominant process of pair-production at the beginning of the shower. From the comparison of the results in Fig. \ref{fig:edep_primaries_energies}, the difference between electromagnetic and hadronic showers will increase with the primary energy.

\subsection{Energy deposition as a function of nature and energy of primary particle}

\begin{figure}
\begin{center}
\includegraphics[width=0.48\textwidth]{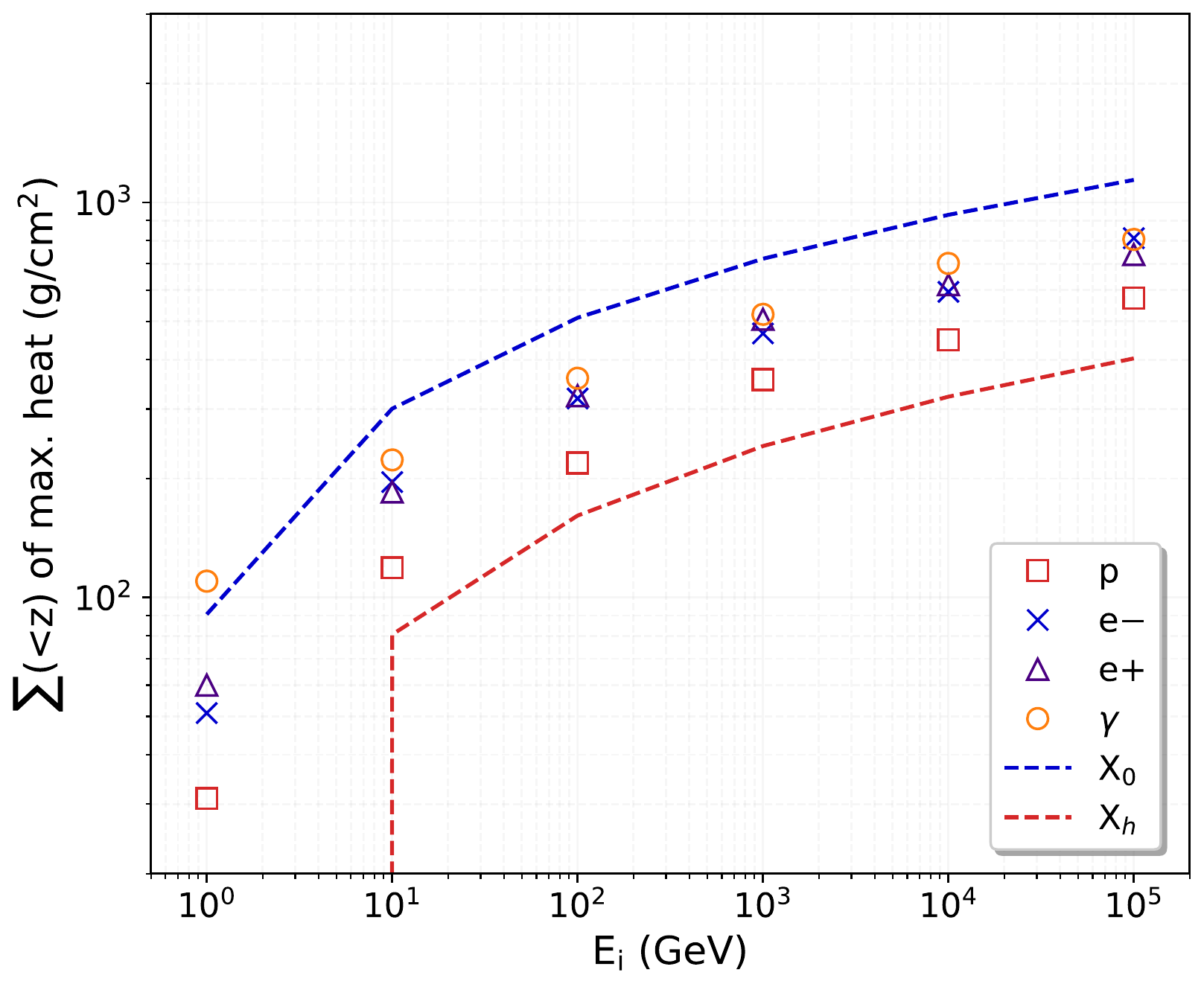} 
\caption{Depth of maximum heating for proton, electron, positron and gamma rays of various initial energies acting as primaries. The dashed lines mark the heating depth for hadronic (X$_h$, red) and electromagnetic (X$_0$, blue) particle showers, respectively, approximated by the Heitler model.}\label{fig:edep_E}
\end{center}
\end{figure}

The results for the heating depths from Fig.~\ref{fig:edep_primaries_energies} are summarized in Fig.~\ref{fig:edep_E}. 
For comparison, we also show the expected heating depth based on the formulas Eq.~(\ref{heat_em}) and Eq.~(\ref{heat_hadr}) derived within the Heitler toymodel for hadronic and electromagnetic particle showers. As expected the analytical values do not predict the exact value for the depth correctly  due to very simplistic assumptions made in the model, but are able to reproduce the general trend and the order of magnitude. Therefore, it is no surprise that the difference between pair and hadron initiated showers is less prominent in the simulations compared to the analytical predictions. The Heitler toymodel is a simplified model to describe general features of particle showers. It doesn't take into account e.g. the dissipation of energy in the form of electromagnetic showers for hadron-induced showers and a constant multiplicity during the shower development.

The values resulting from the simulations differ by roughly an order of magnitude for low- and high-energetic primaries. This means that one should be able to draw conclusions on the initial primary energy from observations of shallow or deep heating in a companion's atmosphere. Whether the determined difference in the values for the single primary types is large enough to distinguish the type of particles in the pulsar wind by observations is questionable and strongly dependent e.g. on the exact knowledge of the companion's atmosphere, even though the discrepancy increases for higher primary energies. On the other hand, the influence of the primary energy should be clearly distinguishable. 

\subsection{Distribution of energy deposition for different atmosphere's composition}\label{sec:composition}
The composition of the atmosphere in which the particle shower is induced could have an impact on the development of the shower and therefore on the deposition of the energy due to the different charge and molar mass of the elements. 

Therefore, we simulate an atmosphere consisting of Helium, with a molar mass of $A=4\,$g/mol and a charge $Z=2$, and an atmosphere consisting of Carbon, with a molar mass of $A=12\,$g/mol and a charge $Z=6$, while keeping the density of $1\,$g/cm$^3$. Pure atmosphere made out of only one element are not realistic, but the purity helps to elaborate the effect caused by the composition.

\begin{figure}
  \centering
    \includegraphics[width=0.48\textwidth]{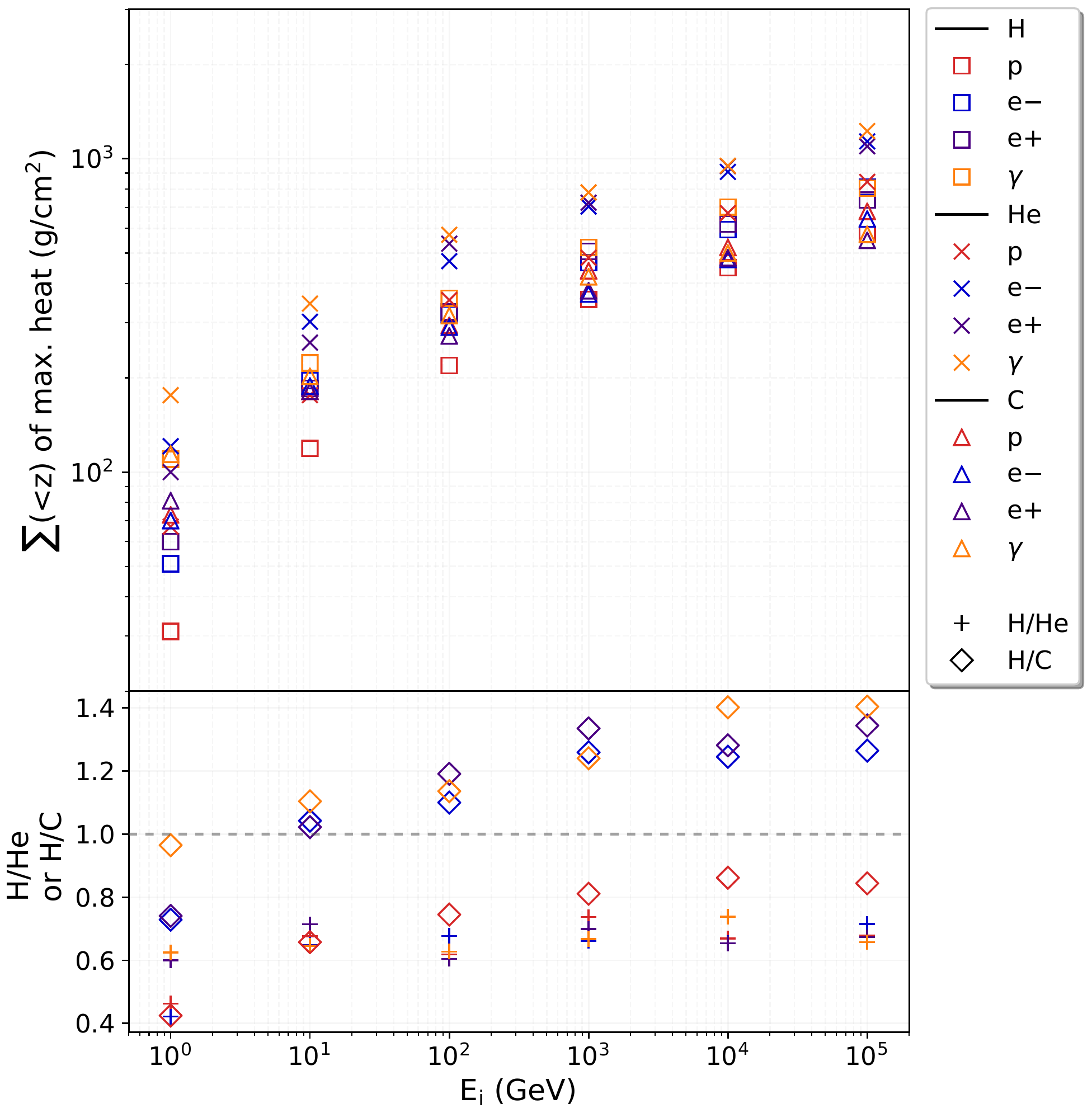}
		\caption{Depth of maximum heating depending on the composition of the companion's atmosphere: pure Hydrogen (H) atmosphere as default compared to Helium (He) and Carbon (C) atmospheres. }
		\label{fig:composition}
\end{figure}

As shown in Fig.~\ref{fig:composition}, in The general trend that leptonic or photonic initiated showers deposit their energy deeper in a Hydrogen atmosphere (squares) than proton-induced showers is also found for the results of a helium (He) atmosphere (crosses) while all primaries lead to a slightly deeper heating in a He-atmosphere. 
For a carbon (C) atmosphere (triangle), the results are different: here, the heating depth for the different primary particle seems to be inverted. Proton-induced showers deposit their energy slightly deeper than photon- or- lepton-induced showers.

\begin{table}
\caption{Radiation length $X_0$ and nuclear interaction length $X_h$ for H, He and C atmospheres with a constant density of $1\,$g/cm$^2$, obtained from \textsc{Geant4}. We neglect the dependency on the energy of the particles in this study for simplicity. }
\begin{center}
		\begin{tabular}{ l | c | c   }
		  atmosphere  &  $X_0$ & $X_h$  \\
		   composition &   (g/cm$^2$) &  (g/cm$^2$) \\ \hline
      H & $63.2\,$ & $35.0\,$   \\	
			He & $94.3\,$  & $55.6\,$  \\		
			C & $42.3\,$  & $80.2\,$   \\		
		\end{tabular}	\label{tab:numbers}
		\end{center}
\end{table}

This behaviour can be also reproduced by the analytical approximations,
 elaborated in Sec.~\ref{sec:particleshowers} and the values for $X_h$ and $X_0$ given in Tab.~\ref{tab:numbers}. For a helium target both values are larger than for hydrogen, leading to a general deeper heating. For Carbon, the radiation length is shorter than for hydrogen which ends in a shallower heating for electromagnetic showers. The nuclear interaction length is longer, leading to a slightly deeper heating for hadron-induced showers.

The difference in the calculated heating depth amounts up to a factor of $2$ for the different compositions of the companion atmosphere. It can thus be seen as a second order effect and we will assume in this work that the composition's impact on the heating depth will not have a significant effect on the heating depth.

\subsection{Effects of density gradient}

We study the impact of the density gradient on the heating depth in the following. Here, we profit from the definition of the transversed column density: the energy deposition depends primarily on interactions with the medium which depend on the transversed column density, while particle decay does not.  As a consequence the density gradient of the atmosphere affects the balance of the two effects and can change the column density at which energy is deposited.

The rate of the energy deposition as a function of the atmospheric depth should differ according to the dominant process (decay or interaction), and thus according to the density profile of the atmosphere.

\begin{figure}
  \centering
    \includegraphics[width=0.48\textwidth]{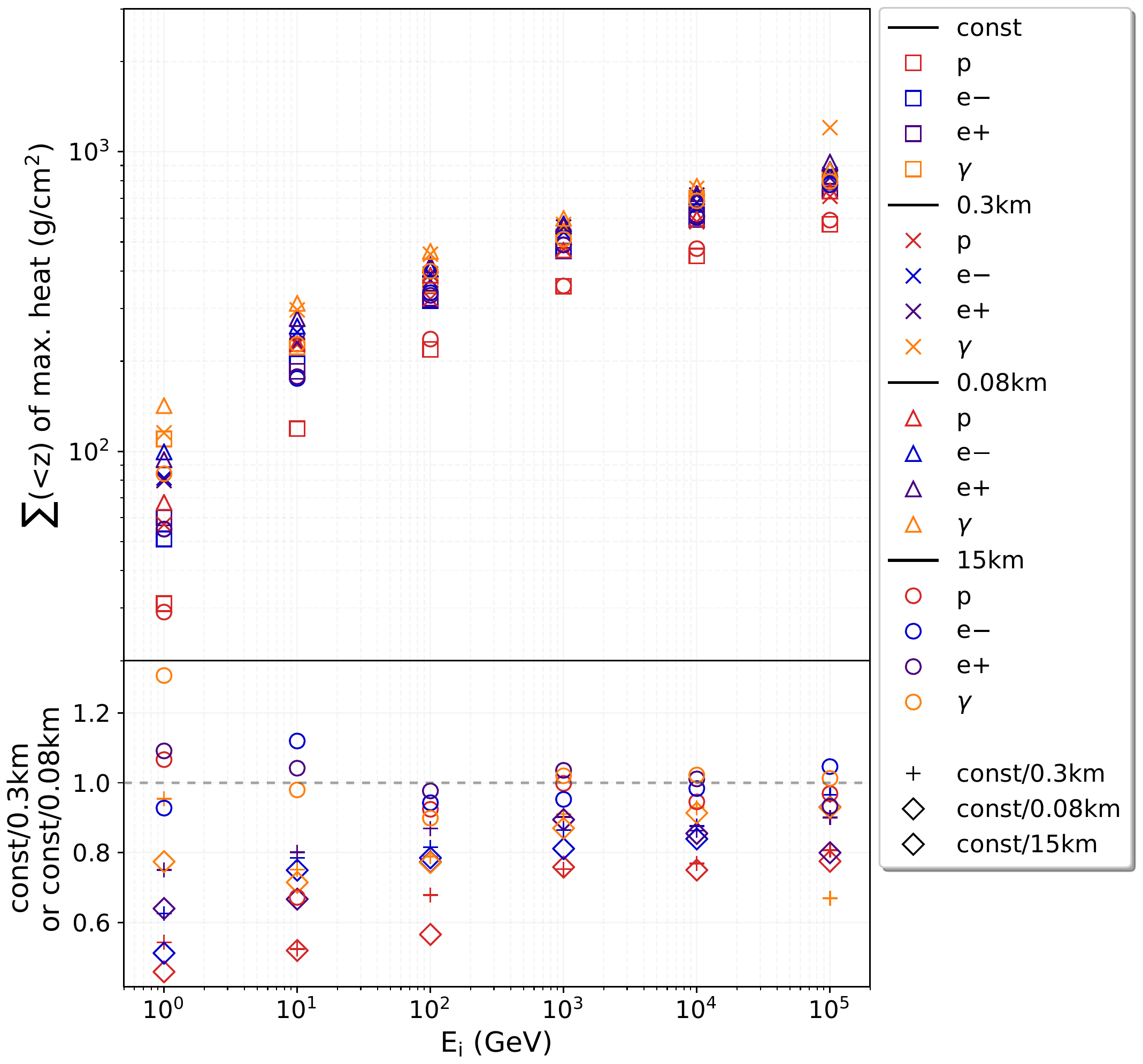}
		\caption{Depth of maximum heating depending on the density gradient in the atmosphere: Results using a constant density profile of the atmosphere with $\rho_0=1\,\textrm{g/cm}^3$ (squares) are compared to density profiles which are described by a exponential function $\rho(z)=\rho_0 \exp(z/h)$ with $h=0.3\,$km (crosses), $h=0.08\,$km (triangles) and $h=15\,$km (circles), respectively, as scaling height.}
		\label{fig:edepdrho}
\end{figure}

In comparison to a hydrogen atmosphere with a constant density of $\rho=1\,\textrm{g/cm}^3$, we simulate in addition atmospheres with density profiles described by an exponential function $\rho(z)=\rho_0 \exp(z/h)$ with scaling heights of $h=0.3\,$km (crosses in Fig.~\ref{fig:edepdrho}), $h=0.08\,$km (triangles) and $h=15\,$km (circles). The gradient is simulated with $N_{\rm layers}=3334$ layers, while layers with no energy deposition will be neglected in the later analysis. Depending on the scaling height, the layers have a width of $150\,$cm and $450\,$cm for the highest scaling height to adjust for density jumps. We adapt the height of the atmosphere according to the given scaling height and numbers of layers so as to guarantee that the induced particle shower is fully contained in the atmosphere.

Figure~\ref{fig:edepdrho} examines the effects of a density gradient on the energy deposition rate. The results for the heating depths do not show any major differences, demonstrating that a detailed knowledge of actual density structure of the companion's atmosphere is negligible for this study.

\subsection{Minimum cut-off energy}
In the simulations, particles can be tracked down to a set energy $E_{\rm cut}$. Below this energy, the particle is considered as ``lost'' in the atmosphere, and their energy is counted as deposited energy. One can follow a particle of each type $i$ down to it's specific energy $E_{{\rm cut},i}$ as done in traditional Monte Carlo simulation. 
But this may lead to an imprecise stopping location, and in addition it would be particle's type and material dependent. Therefore, \textsc{Geant4} uses instead tracking cuts for gamma, electron, positron and proton, by introducing a range cut\footnote{\url{http://geant4-userdoc.web.cern.ch/geant4-userdoc/UsersGuides/ForApplicationDeveloper/BackupVersions/V10.4/html/index.html}}. When a particle has no longer enough energy to produce secondaries which travel at least this distance, the discrete energy loss stops while the particle is tracked down to zero energy using continuous energy loss. The chosen tracking threshold, given as a range, is internally converted to an energy threshold which depends on the particle type and material.

Setting a low cut-off energy can slow down the code considerably, since the number of produced secondaries for high-energetic primaries is huge. Therefore, we chose a range cut of $6\,$cm as a default value for the simulations of the presented study.

The default value for the presented study is $6\,$cm, while the default value in \textsc{Geant4} is set to $1\,$mm. For comparison, we also investigated a range cut of $1\,$cm. The corresponding energy thresholds, determined by \textsc{Geant4}, are given in Tab.~\ref{tab:cutsenergy}.

\begin{table}
\caption{The corresponding energy thresholds for set range cuts in an Hydrogen atmosphere. The thresholds depend also on the particle type. Values are obtained from \textsc{Geant4} .} 
\begin{center}
		\begin{tabular}{ c | c | c | c | c | c | c  }
		  range cut &  $\gamma$ & $e^-$ & $e^+$ & $p$  \\ \hline
      $1\,$mm & $990\,$eV & $586\,$keV & $570\,$keV & $100\,$keV \\ 	
			$1\,$cm & $990\,$eV & $4.23\,$MeV & $4.00\,$MeV & $1.00\,$MeV \\ 	
			$6\,$cm & $990\,$eV & $27.8\,$MeV & $26.3\,$MeV & $6.00\,$MeV \\ 
		\end{tabular}	\label{tab:cutsenergy}
		\end{center}
\end{table}

\begin{figure}
  \centering
    \includegraphics[width=0.48\textwidth]{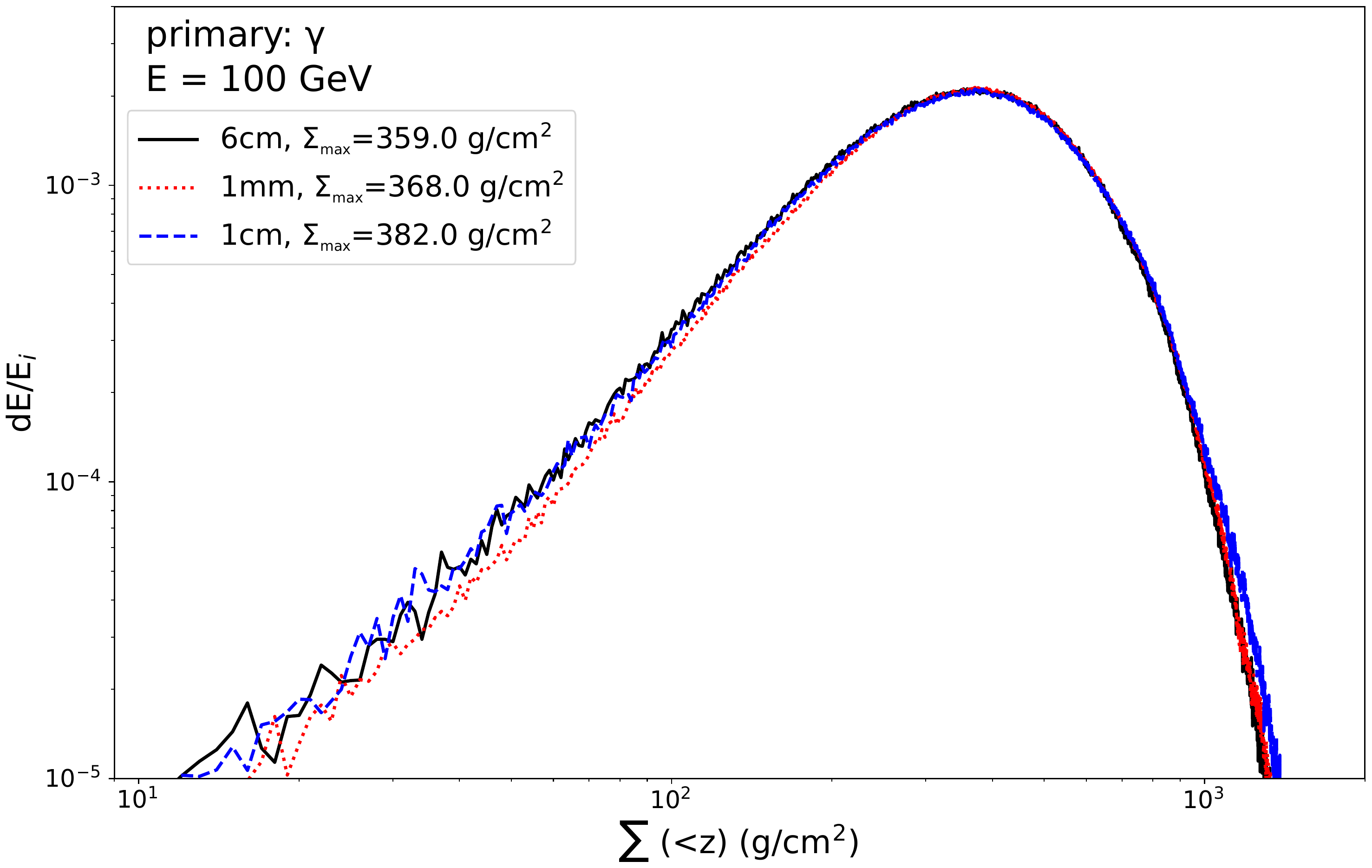}
		\caption{Energy deposition 
		as function of the transversed density: range cut of $6\,$cm as default for the following simulation study compared to $1\,$mm ({\sc Geant4} default) and $1\,$cm: results for a gamma ray primary of the energy of $100\,$GeV. }
		\label{fig:cuts_single}
\end{figure}

\begin{figure}
  \centering
    \includegraphics[width=0.48\textwidth]{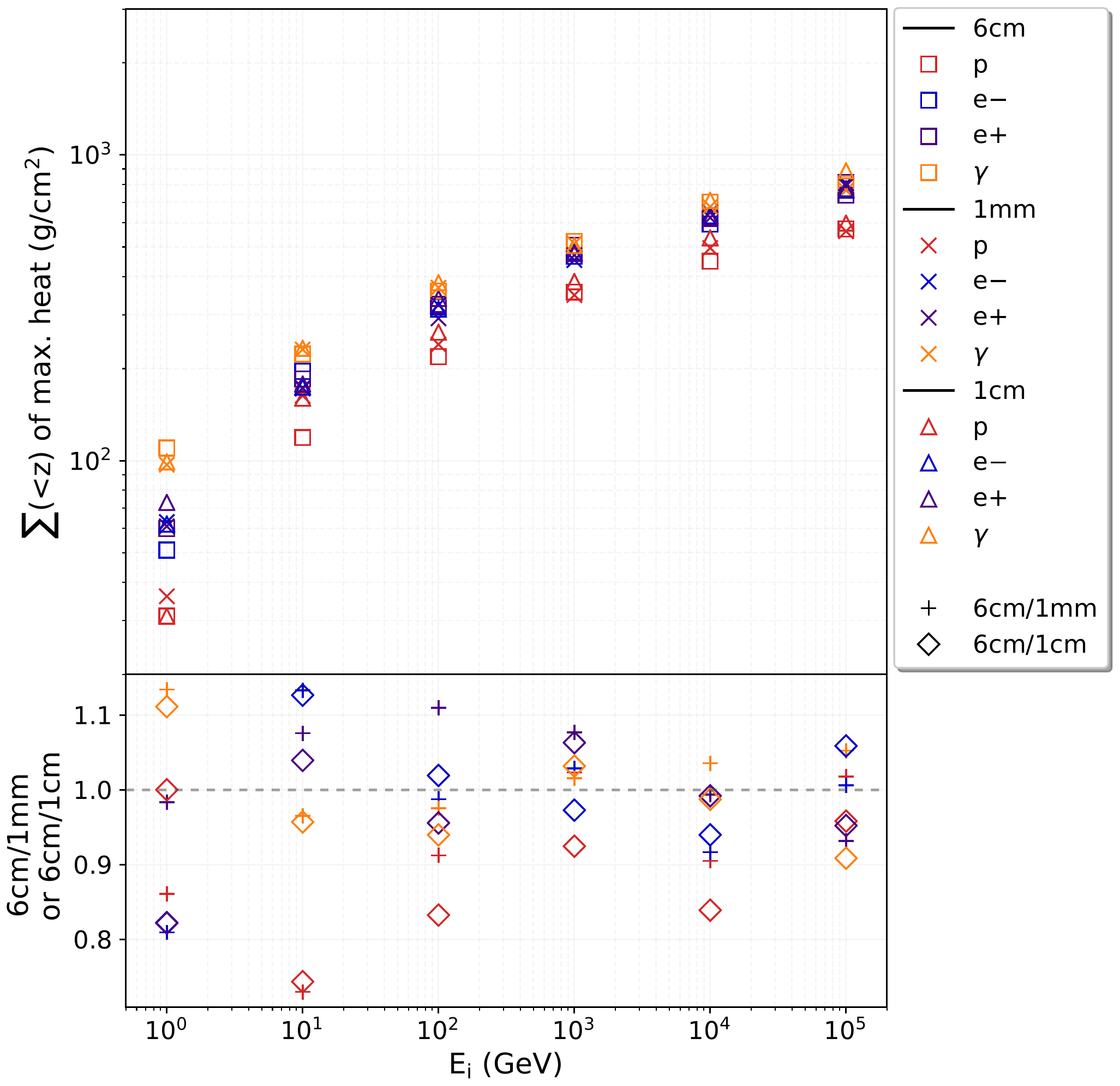}
		\caption{Depth of maximum heating depending on the used range cut: $6\,$cm as default compared to $1\,$mm (\textsc{Geant4} default) and $1\,$cm. }
		\label{fig:cuts_summary}
\end{figure}

Figure~\ref{fig:cuts_single}, displaying the results for a gamma-ray primary of an energy of $100\,$GeV, demonstrates that the energy deposition does not vary strikingly for different included range-cut values. This is also valid for all studied primaries and energies, as shown in Fig~\ref{fig:cuts_summary}.

\subsection{Dependence on interaction models}
For the presented study we used Physics Lists (see Sec.~\ref{sec:setup}) provided by \textsc{Geant4} as key reference physics lists: QGSP\_BERT, FTFP\_BERT and QBBC.
All studied physics list use the "standard" \textsc{Geant4} electromagnetic physics as built by the \textit{G4EmStandardPhysics} constructor (EMOpt0). 
It handles all the processes relevant for for $\gamma$, $e^-$, $e^+$, $\mu^-$, $\mu^+$, $\tau^-$, $\tau^+$-particles and all stable charged hadrons/ion. For more details, see the documentation of the EM physics constructors\footnote{\url{http://geant4-userdoc.web.cern.ch/geant4-userdoc/UsersGuides/PhysicsListGuide/html/electromagnetic/}}. The main difference consists in the treatment of the hadronic interaction. The documentation of the reference lists can be found here\footnote{\url{http://geant4-userdoc.web.cern.ch/geant4-userdoc/UsersGuides/PhysicsListGuide/html/physicslistguide.html}}. We chose QGSP\_BERT as the default reference list for the study

\begin{figure}
  \centering
    \includegraphics[width=0.48\textwidth]{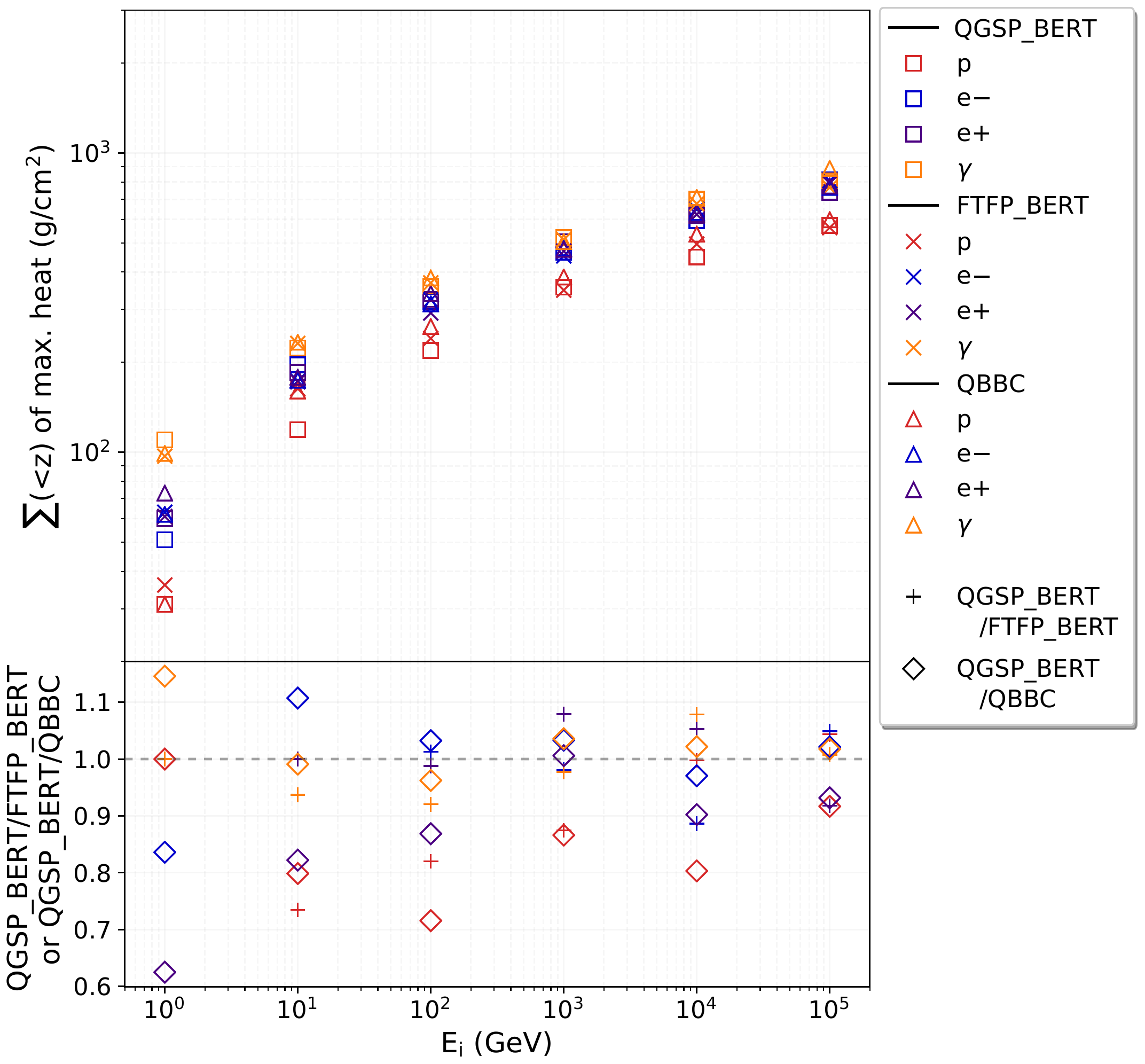}
		\caption{Depth of maximum heating depending on the used interaction: results for QGSP\_BERT as the default physics list compared to FFTP\_BERT and QBBC, with a constant Hyrdogen atmosphere, and a range cut of $6\,$cm. }
		\label{fig:interaction_models}
\end{figure}

Results for the calculated heating depth on the basis of the different Physics List are shown in Fig.~\ref{fig:interaction_models}. There is no significant impact by the chosen Physics List observable. Therefore, the actual choice of the interaction model seems to have a negligible effect on the outcome of the study.

\subsection{Effects of companion magnetosphere}\label{section:magnetosphere}

Deflections in the companion magnetic field could affect the propagation of the lowest energy particles in the atmosphere. The magnetic field of any spider companion is so far unknown. Some models indicate that the surface magnetic field strength of companions could as high as $B_{\rm c}\sim 10^4\,$G (e.g., \citealp{Romani15,Romani16,Kao18}, for extreme systems). Considering that $10^4$\,G is particularly high for main sequence stars, we will use in the following an estimate of $B_{\rm c}=10^3\,$G, which could still be seen as the conservatively high.  
Assuming a dipole magnetic field, at a radius $r_{\rm atm}$\footnote{Not to be confused with the atmosphere scale height $h$. of order $R_{\rm c}$,} an impinging charged particle of energy $E$ would have a Larmor radius 
\begin{eqnarray}\label{eq:rL}
r_{\rm L}&\sim& 0.03\,{\rm km}\,\frac{E}{1\,{\rm GeV}}\left(\frac{B_{\rm c}}{10^3\,{\rm G}}\right)^{-1} \left(\frac{r_{\rm atm}}{R_{\rm c}} \right)^{3} \ . 
\end{eqnarray} 
For energies $E\gtrsim 1\,$GeV, the particle Larmor radius is much greater than the penetration depth of the particle (of $\sim 100\,$cm). 
Therefore, particle deflection in the atmosphere magnetic field may be neglected.

This is confirmed numerically, by simulations run with homogeneous magnetic fields  parallel or perpendicular to the air-shower direction (the real-life magnetosphere structure will be a mixture of those). Parallel (i.e., radial in the companion point of view) fields, as would be found at the stellar poles, could enable to reach penetration depths as deep as without magnetic fields. However, for stronger magnetic field strength or highly turbulent magnetic fields, particles could be affected.

On the other hand, Eq.~(\ref{eq:rL}) shows that particles with $E\lesssim 100\,$TeV will be deflected over scales $r_{\rm L}\lesssim R_{\rm c}\sim$ few thousand kilometers. It is not guaranteed that these particles can reach the top of the companion atmosphere (located at radius $\sim R_{\rm c}$), due to magnetic deflection. This simple estimate constrains the energy of the impinging charged particles to $\gtrsim 100\,$TeV. Below this energy, only photons would be unaffected by the magnetic field and could reach the atmosphere. These considerations however depend on the structure of the magnetic field. Charged particles below 100\,TeV could leak in from the poles or other favorable structures.

\section{Application to benchmark binary systems}\label{section:application}

In this section, we compute atmosphere models for a set of companion parameters (see Fig.~\ref{fig:application}), as well as for existing black-widow and redback systems (see Fig.~\ref{fig:application_examples}), and compare their column densities to the heating depths calculated in the previous sections. This enables us to assess the energy of particles in the wind at the distance of the companion.

\subsection{Numerical modeling of companion atmospheres}\label{subsection:atm_model}
Low-mass helium white dwarfs are the most frequent MSP companions \citep{Tauris12}, which is expected as a result of the standard recycling scenario. The majority of observed helium white dwarfs shows signatures of pure hydrogen atmospheres, more abundant than helium or other metals. Due to gravitational settling, 
helium and heavier elements are effectively removed from the atmosphere toward inner layers \citep[and references therein]{Rohrmann:2001ku}. 
Mixing by a strongly convective flux could lead to a helium-rich atmosphere in cool white dwarfs, as observed for example for PSR J0740+6620 \citep{Beronya:2019lwc}, mainly in the inner layers of the white dwarf atmospheres, in particular in the zones of the atmosphere where hydrogen is partially ionized or dissociated. This yields a significant drop in the value of the adiabatic temperature gradient which favours the convective instability.

A set of models for atmosphere structures for low-mass helium white dwarfs with pure hydrogen atmospheres,  shown in Fig.~\ref{fig:application}, were generated with the numerical code described in \cite{Rohrmann:2001ku,2002MNRAS.335..499R}. This code provides non-gray models for pure-hydrogen, pure-helium, or mixed H/He atmospheres under the assumptions of plane-parallel geometry (surface gravity is constant through the atmosphere), hydrostatic equilibrium, local thermodynamic equilibrium for the gas and constant total energy flow. The energy flow is due to electromagnetic radiation and convection (whenever the Schwarzschild criterion for convective instability is verified). Radiative transfer is rigorously calculated solving transport equations with detailed gas opacities. The model incorporates the occupational probability formalism of \cite{1988ApJ...331..794H} to treat non-ideal effects in the gas equation of state and
 the opacity in high-lying atomic levels. The models include a complete description of hydrogen line and edge opacities, collision-induced absorption opacity for both hydrogen molecules and helium, and are flux calibrated to Vega zero colors. 

Physical quantities throughout the whole atmosphere, such as temperature, density, and pressure, are evaluated verifying all equilibrium conditions. Consequently, calculated emergent spectra and other associated results (magnitudes, colors) correspond to homogeneous, stationary, non-irradiated atmospheres. 

The gas density in the atmospheric layers has a strong dependence on the gravity force through the hydrostatic equilibrium law. For most stars, the surface gravity $g$ is usually determined by  fitting the absorption lines in the spectrum, since the line broadening processes (involving particle collisions/interactions) depend on the gas density and temperature. However for stars with low quality spectral observations or without lines in their spectra (as it is the case of very cool white dwarfs), this method cannot successfully constrain $g$. In such case, $g$ may be determined from the stellar mass $M$ and the distance from the Earth, for example by fitting the stellar brightness or the apparent bolometric magnitude using atmosphere models. This method allows us to evaluate the stellar radius $R$ and then calculate $g=GM/R^2$.

In practice, one does not usually have the bolometric magnitude (that involve the radiative flux over the whole spectrum), but magnitudes in a set of instrumental filters which measure partial energy fluxes in different parts of the energy spectrum. For some pulsar companions, $\log g$ is poorly constrained in this way as well, due to difficulties in finding a unique model that fits all data sets. Indeed, companions of millisecond pulsars, especially black widows such as PSR B1957+20 (Section~\ref{subsection:B1957}) with large Roche-lobe filling factors, can have strongly distorted atmospheres, and their observational magnitudes may show deviations with respect to typical homogeneous atmospheres of isolated and non-rotating stars \citep{2000A&A...364..265O,Reynolds07}.

For the present work however, it suffices to have rough estimates of $g$, and approximate atmosphere models which provide averaged density profiles.

\subsection{Inferring particle energy the pulsar winds}
\begin{figure}
\centering
\includegraphics[width=0.48\textwidth]{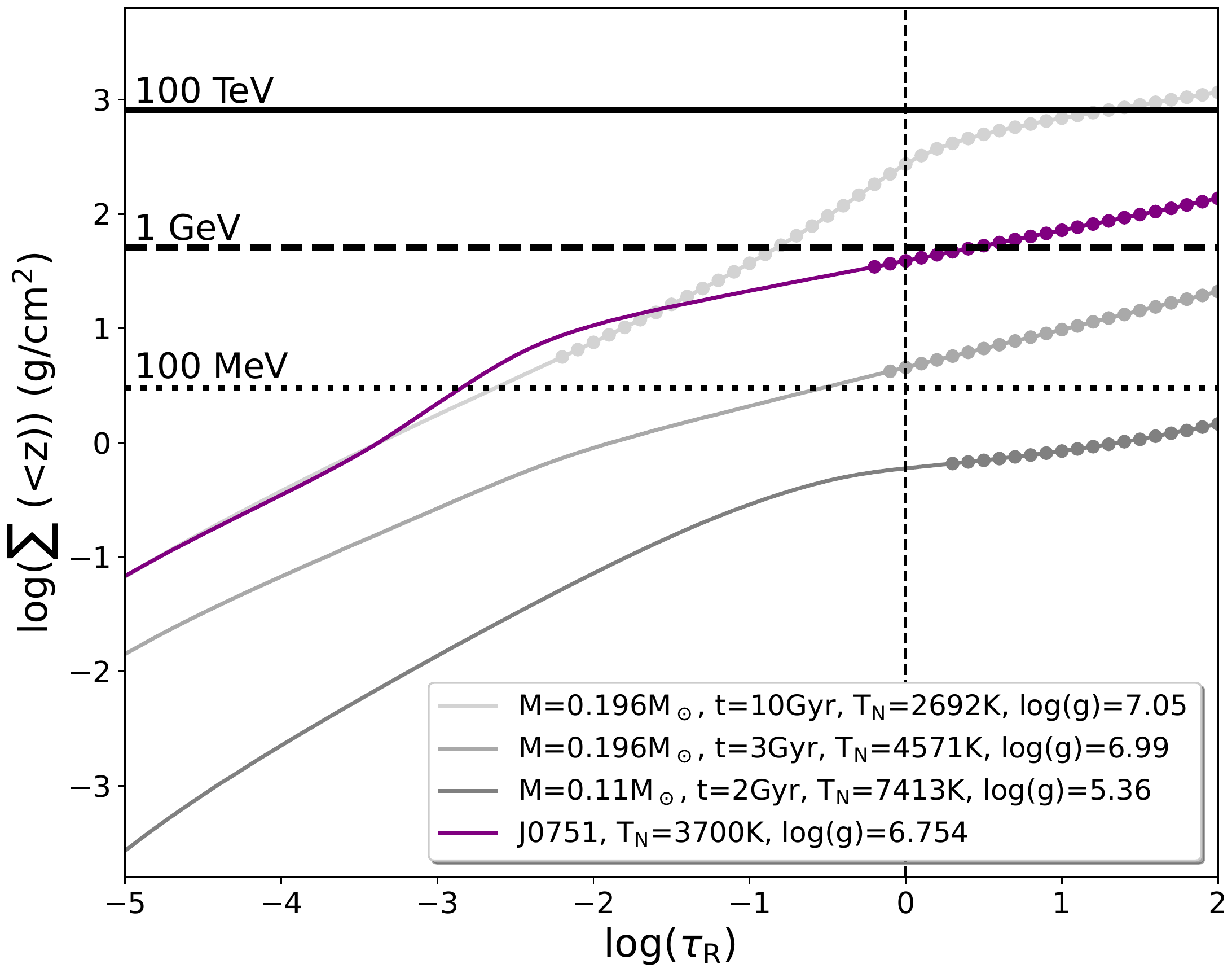}
\caption{
Hydrogen atmospheres for three low-mass He-core white dwarf models, with parameters as labelled above the plot (gray lines) and one MSP-WD binary J0751+1807 (purple line) . Transversed column density in the atmosphere down to $z$, as a function of the Rosseland-mean optical depth at $z$. The photosphere is located at $\log \tau_R \sim 0$ and indicated by the black vertical dashed line. Depths where more than half the flux is carried by convection are indicated by small circles. 
The horizontal lines represent the expected maximum heating depth $\Sigma_{\rm max}$ for injected primary electrons of different energies. 
The first horizontal line from the top (solid) indicates the column density for which a shower resulting from a $100\,$TeV electron (impinging vertically from above the atmosphere) reaches its maximum heating. 
The next horizontal lines (second from top, dashed) indicates respectively the results for the maximum heating for an electron-induced shower of $1\,$GeV, and the bottom line (dashed) for a shower of $100\,$MeV. 
}
\label{fig:application}
\end{figure}

Figure~\ref{fig:application} shows the logarithm of the transversed column density as a function of the Rosseland mean optical depth $\tau_\mathrm{R}$ which describes the average optical depth in a gas. A value of $\tau_\mathrm{R}=1$ marks the mean position of the photosphere, represented by a vertical dashed line. Above this layer ($\tau_\mathrm{R}<1$) the atmosphere is transparent for photons, below it is opaque.

If the heating by a particle shower takes place above the photosphere, the temperature of the atmosphere will not change. But a shallow heating on the top of photosphere can produce an inverted temperature profile and consequently an emission in the core of some spectral lines. This is due to the fact that line cores are more opaque (higher radiative cross section) than line wings and are therefore formed in more superficial layers. However the detection of core emission depends on the resolution and signal-to-noise ratio of the observations. For the present instrumental capacities, these heating signs (line core emission) could remain undetectable in optical companions of pulsar systems. 

If the heating depth is below the photosphere, the observed temperature of the companions atmosphere will increase, still showing a usual photospheric temperature profile. Here, the observable absorption lines will be shifted to the ``new" temperature.

Deep in the atmosphere, convection dominates over radiation for energy transfer. In general, it seems unlikely that an external energy irradiation could be redistributed over the whole surface by convection to completely wash out the day/night temperature variations. For such a process to take place, it is necessary to invoke a special greenhouse-type effect (layers at the top of the atmosphere with high opacity values for outcoming energy flux) so that the transverse convection energy flux becomes more efficient that the vertical one. 

In some systems, deep heating could however affect the convective stability and the atmospheric structure of the star. The heating shuts off convection on the irradiated side; the change in the outer boundary condition induces the heat carried up by convection to be trapped on that side, changing the structure to deeper and deeper layers until winds carry heat to the night side. Such a process could lead to a uniformisation of the temperature over the stellar surface (Jermyn \& Phinney in prep.).\\

The gray lines in Fig.~\ref{fig:application} show the evolution of the optical depth dependent on the transversed column density for several models for white-dwarf atmospheres. Here, the circles mark the location in the atmosphere which are dominated by convection. The characteristic parameters for each atmosphere model are given in the label above, ordered from the top to the bottom. The scale heights for white dwarf atmosphere 
$h$ for the shown models are $0.08\,$km, $0.36\,$km and $15\,$km for a temperature of $2692\,$K, $4571\,$K and $7413\,$K, respectively, representing night-side temperatures. 

Fig.~\ref{fig:application} shows also the calculated heating depth for three monoenergetic electron  beams, acting as examples, with energies: $100\,$TeV, $1\,$GeV, and $100\,$MeV. We saw in the previous section that the nature of the particles (protons, electrons or else) does not have a major impact on the heating depth. We thus chose to show one type of particle here for simplicity.

In cool atmospheres ($T_{\rm night}\lesssim 3000\,$K), where the opacity is dominated by induced dipole H$_2$ interactions and non-ideal effects, showers with energy $E\lesssim 100\,$TeV deposit their energy near or above the photosphere. In atmospheres with higher temperatures ($\gtrsim 5000\,$K) most showers, down to $\sim 100\,$MeV primaries deposit their heat well below the photosphere. Companions showing temperature variations, which should thus undergo deep heating, provide a lower limit on the particle energy in the wind. This lower limit is all the more constraining with higher values, as the companion has a cooler night side. 

\begin{figure}
\centering
\includegraphics[width=0.48\textwidth]{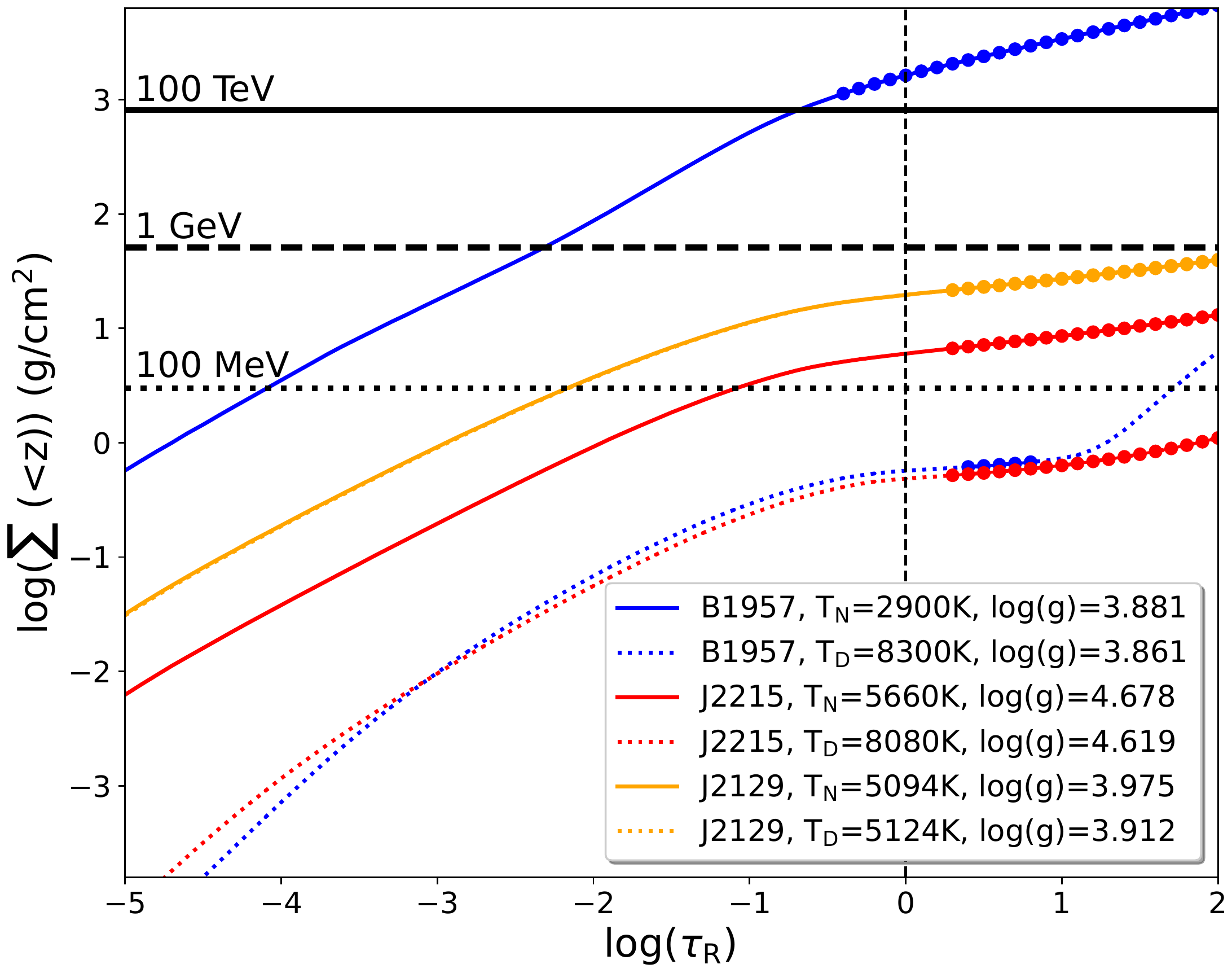}
\caption{The results for the simulated companion atmospheres of PSR B1957+2048J (blue), J2129-0429 (orange) and J2215+5135 (red). The solid lines represent the atmospheres obtained for the night-side temperatures, the dotted lines the ones for the day-side, respectively. For J2215, due to the tiny differences for the day and night sides, the lines overlap.
The horizontal lines represent the expected maximum heating depth $\Sigma_{\rm max}$ for injected primary electrons of different energies, as shown in Fig.~\ref{fig:application}. }
\label{fig:application_examples}
\end{figure}

Note that the comparisons with model atmospheres are made with unperturbed atmospheres: the feedback from heat-deposition is not taken into account. Our simulation results depend mainly on grammage, hence this effect is unlikely to impact the heating depth. However, the photospheric depth of the companion could change. In order to illustrate the difference in photospheric depth, we plot in Figure~\ref{fig:application_examples} atmospheres obtained with both $T_{\rm night}$ and $T_{\rm day}$. A study using the former model provides constraints on the nature of a pulsar wind that impinges a cool companion atmosphere, which could have been the case in the first evolutionary stages of the system, assuming that the companion had initially similar characteristics and homogeneous $T_{\rm night}$ temperature. The latter model rather answers the following question: ``What type of pulsar irradiation enables to maintain the companion atmosphere at its observed heated temperature?", and should be less constraining in terms of energy deposition.

For companions with uniform temperature, shallow heating could be operating, in which case more accurate observations could reveal inverted temperature profiles and absorption/emission lines. This can be translated into an upper limit on the wind particle energy. Depending on the atmosphere model (for hot systems in particular), the upper limit to the particle energy can be surprisingly low ($\lesssim 100\,$MeV). An explanation implying that the pulsar wind is still Poynting flux dominated cannot stand, as stellar-size objects should also be heated up by Poynting flux at this distance \citep{Kotera2016}. An alternative, natural, explanation could be that the companion is not or no longer tidally locked to the pulsar. Radiation times $t_{\rm rad}$ longer than the spin period are then necessary to sustain this interpretation. Finally, another possibility is that deep heating is occurring in ranges where convection can operate efficiently to uniformize the temperature over the stellar surface. This requires specific atmosphere structures and that the convection time be shorter than the radiation time.

We discuss in the following the interpretation derived from our models for four particular binary systems (see Fig.~\ref{fig:application_examples}).

\subsection{The prototype heated black widow: PSR B1957+20} \label{subsection:B1957}
PSR 1957+20 is the original and one of the best studied members of its class (e.g., \citealp{Phinney88,Kluzniak88,Arons93,Callanan95,Khechinashvili00,Stappers03,Reynolds07,vanKerkwijk11,Breton:2013ess}). It has been subject to many observations in many energy bands (from radio to X-rays and $\gamma$-rays), and to many theoretical interpretations. PSR B1957+20 consists of a 1.6 ms radio pulsar orbiting a companion of mass no less than 0.022 $M_\odot$, in a binary of orbital period 9.17\,h, and can be therefore classified as black widow. For 10 per cent of this orbit, the radio emission from the pulsar is eclipsed \citep{1988Natur.333..237F}. \cite{Reynolds07} find that the effective temperature of the companion varies from $T_{\rm day}\sim 8300$~K to $T_{\rm night}\sim 2900$~K between the day/night sides. The corresponding irradiation temperature yields $T_{\rm irr} = 8269\,$K, and the efficiency and irradiation ratios $f_e=75.5$ and $\eta_{\rm irr}=87\%$. The companion is expected to be highly irradiated and shows an observable heating effect.

As briefly discussed in Section~\ref{subsection:atm_model}, due to inhomogeneous physics conditions of its perturbed atmosphere, it is not possible to reproduce the magnitudes of B1957+20 as measured by \citealp{Reynolds07} with a single model. Therefore, in order to obtain constrains over the value of $g$ for PSR B1957, we adopted the most reliable results of previous studies, instead of trying to adjust photometric data with unperturbed atmosphere models. 
With the updated values of \cite{vanKerkwijk11} (pulsar mass $M=2.4\,M_\odot$, companion mass $M_{\rm C}=0.02\,M_\odot$) and the orbital period ($P_{\rm b}=9.1672\,$h), one finds an effective Roche-lobe radius of $R_{\rm L}=200\,000\,$km. This implies $R=0.8\, R_{\rm L}\sim 160\,000\,$km $\sim 0.23\,R_{\odot}$. 

Figure~\ref{fig:application_examples} presents the column density evolution of an atmosphere model with $\log g = 3.881$ and $T_{\rm night}=2900\,$K (solid blue) and $\log g = 3.861$ and $T_{\rm day}=8300\,$K (thin dotted blue), which mimics respectively the non-illuminated and illuminated sides of the B1957+20 companion. The values for the local gravity accelerations $g$ were calculated from the Roche model. They take into account the binary dynamics and respective orbital parameters of the system. 
The heating depth has to occur below the photosphere, constraining the wind particle energy to $\gtrsim 100\,$TeV at distances from the pulsar of $a=2.5\,R_\odot$, using the cool atmosphere model. 
The hot atmosphere model is not constraining in terms of particle energy, and can be maintained with particles with energy $< 100\,$MeV.

\subsection{A heated redback: PSR J2215+5135} 
PSR J2215+5135 is a well studied representative of the group of redback binaries, with a period of 2.6 ms and a companions mass of 0.33 $M_\odot$ in a orbital period of 4.14 h, the shortest period among the Galactic field redbacks~\citep{Breton:2013ess, Linares:2018ppq}. 
As typically observed for redbacks, the night-side temperature ($T_{\rm night}\sim$ 5660 K, inferred from spectra) is hotter than compared to black-widow companions. The temperature of the day side is determined to $T_{\rm day}\sim 8080\,$K which indicates a strong heating effect by the impinging pulsar wind. 
The corresponding irradiation temperature yields $T_{\rm irr} = 7542\,$K, and the efficiency and irradiation ratios $f_e=6.68$ and $\eta_{\rm irr}=47\%$\footnote{We note here the inconsistency with the value given in~\cite{Linares:2018ppq}.} following Eq.~\ref{eq:flux} and \ref{eq:radeffiency} and parameters from Tab.~\ref{tab:pulsars}, expressing that the expected strong heating is observed in this system.
Furthermore, the companion's atmosphere shows strong absorption lines, metallic on the night side and Balmer lines on the day side. Their effects on radial velocity measurements of the system are studied in detail with photometric and spectroscopic data in~\cite{Linares:2018ppq}.
The radio dispersion measure gives a distance of about $d=2.9\,$kpc \citep{2013ApJS..208...17A}, and \cite{2014ApJ...793...78S} provide $B, V, R$ magnitudes in different phases.

We show in Figure~\ref{fig:application_examples} the atmosphere model for the day and the night sides corresponding to our closest fit to the photometry of J2215. The system has a companion closely filling its Roche lobe with a filling factor of $0.95$ and a Roche radius of $0.36\,R_\odot$. The values for $\log g \sim 4.6$ for the day and the night side are calculated from the Roche model, taking into account the binary dynamics and respective orbital parameters of this system.

The clear temperature modulation indicates that heat should be deposited below the photosphere. The initial (night-side) temperature of the companion being rather elevated, the photosphere is located at shallow column density. The deep heating constraint only provides a mild lower limit on the energy of particles in the wind above $100\,$MeV at distance $a\sim 1.53\,R_\odot$ from the pulsar.  
For the day side, the system is even less constraining than the B1957+20 system. The temperature of the atmosphere can be maintained with particle energies of $< 100\,$MeV.

\subsection{A redback with mild temperature modulation: J2129-0429}

PSR J2129-0429 has a non-degenerate companion of mass $M_{\rm C}=0.44\,M_\odot$, which is 95\% Roche-lobe filling \citep{Bellm:2015dfa}. It is seen as a system early in its recycling phase with extreme parameters. In terms of atmosphere modeling however, this system seems to be less controversial than the previously analyzed ones. The model presented in Fig.~\ref{fig:application} closely reproduces the observed $R$ magnitude (16.4 against 16.5).

Although the irradiation temperature $T_{\rm irr}^4 = T_{\rm day}^4 - T_{\rm night}^4\sim 2000\,$K is well constrained, $T_{\rm night}$ and $T_{\rm day}$ are equal within error bars as a result of the large correlation between these two parameters.

Due to the large orbital separation  ($P_{\rm b}= 15.2\,$h, $a=3.9\,R_\odot$) the irradiation by the pulsar is unlikely to be efficient which could be the favored explanation to the observation of the weak day-night effect. The quoted irradiation efficiency is  $\eta_{\rm irr}=3\%$ \citep{Bellm:2015dfa} and the corresponding efficiency $f_{\rm e}\sim 10$. This supports the mild temperature modulation observed in this system.

Furthermore, it can be assumed that the companion may not be tidally locked~\citep{Bellm:2015dfa}; this is supported by the fact that the observations are consistent with a value of $0.5$ for the corotation (rotation at half of the orbital speed). On the other hand, the thermal time scale $t_{\rm rad}\sim 5\, {\rm min}\,(T_{\rm eff}/3700)^{-3}(\Sigma_{\rm heat}/10\,{\rm g\,cm^{-2}})$ in the case of deep heating is rather short.

However, it can be noted that, for redbacks in general, even a strong irradiation does not usually lead to a large temperature differences, as the base temperature of the star is much larger than for black widows, and brightness increases with the 4th power of the temperature. 

To account for the observed irradiation temperature, some heat should be deposited below the photosphere, but Figure~\ref{fig:application_examples} shows that the deep heating constraint does not provide a stringent lower limit on the energy of particles in the wind. The particle energy should be above few $100\,$MeV at large orbital separation of $a=3.9\,R_\odot$.

If a shallow heating were to be invoked, the energy of particles in the wind would be constrained to below a few hundred MeV, which is a stringent constraint. This interpretation is not necessarily favored given that the source was first discovered by the {\it Fermi}-LAT \citep{2015ApJS..218...23A}, and should hence accelerate leptons to energies at least higher than few GeV.

\subsection{A white-dwarf companion with no temperature modulation: PSR J0751+1807}

PSR J0751+1807 is a low-mass millisecond pulsar of period 3.48 ms and mass $M=1.26\,M_\odot$ \citep{Nice2008}, having a white dwarf companion of mass 0.12 $M_\odot$, a short orbital period of 6.3 h and only $6\%$ Roche-lobe filling. It is located at a distance $d=400\,$pc \citep{2012ApJ...755...39V}, a statistically corrected distance compared to the ones  published by \cite{2005ApJ...634.1242N,Bassa06}. Given the expected irradiative flux of the pulsar wind incident on the companion, one would expect the presence of a day/night variation.  The corresponding efficiency results in $f_e=3.6$. But surprisingly a modulation in temperature (about $3700\,$K for the day and the night side) could not be observed.

The atmosphere of the companion is not straightforward to model satisfactorily. The fits provided by \cite{Bassa06}, which indicate a pure helium atmosphere, or a helium atmosphere with some hydrogen mixed in, 
 do not include the latest gas opacities concerning collision-induced absorptions (CIA). Their conclusions are thus not validated by the current atmosphere models. We discuss the modeling of this atmosphere in more details in Appendix~\ref{app:J0751}.

In Figure~\ref{fig:application}, we choose to present an atmosphere with parameters $\log g=6.754$ (obtained using $M=0.12\,M_\odot$ and $d=400\,$pc and the Roche model) and $T_{\rm night}=3700\,$K. As can be seen in Fig.~\ref{fig:J0751} simple black body with such parameters may fit the observations, although this implies that the opacities of H and He atmospheres are not yet fully understood in cool temperatures and high densities. 
As demonstracted in Sec.~\ref{sec:composition}, the actual composition of the atmosphere has minor effects on the heating depth.

One possible interpretation of the absence of temperature modulation in this system is the line of \cite{Bassa06}. Initially the pulsar companion was tidally locked. After the end of the mass transfer, the companion contracted to a white dwarf, spun up to conserve the momenta to a rotation period of $5-20$ minutes, much shorter than the estimated thermal time scale  $t_{\rm rad}\sim 60\, {\rm min}\,(T_{\rm night}/3700)^{-3}(\Sigma_{\rm heat}/600\,{\rm g\,cm^{-2}})$. Here, the illustrative depth of 600\,g/cm$^{-2}$ corresponds roughly to the deep heating limit read on Fig.~\ref{fig:application} for this system.
Hence the rotation is so rapid that temperature modulation is erased, just as the temperature of the Earth’s upper atmosphere does not vary much between day and night, since the thermal time is longer than a day. 

It is also possible that shallow heating is operating, in which case the energy of particles in the wind would be constrained to about a few GeV. In that case, spectral analysis should show specific signatures, that have not been observed so far.

\section{Conclusion and Outlook}\label{section:conclusion}
Binary systems such as  millisecond pulsars with white dwarf companions or ``black widows" and ``redback pulsars" provide a unique opportunity to test the nature of the pulsar wind. 
Under the assumption that a non-negligible fraction of the pulsar rotational energy impinges the companion atmosphere under the form of high-energy photons or particles, the atmosphere of the companion is heated and shows a strong day/night variation in the temperature. 
In the presented study we assume that the heating is caused by high-energetic particles inducing a particle cascade in the atmosphere. During the development of the cascade, the complete primary energy will be deposited in the atmosphere.
Depending on depth of the maximum energy deposition, it could lead to observable changes in the temperature profile of the atmosphere, as for instance an increase of the  temperature or day-night effects.

To study the effects of the heating by an impinging particles of a pulsar wind, we performed \textsc{Geant4} simulations of the shower development in simplified atmospheres. We could show that neither the exact density profile of the companion atmosphere nor the exact composition has to be known to calculate the depth of the maximum heating. 

We conclude from our simulations that the observation of heating effects can constrain the energy of the incident particles of the pulsar wind. On the other hand, the nature of the primary particle (photon, electron or proton) will be hardly distinguishable --which increases the robustness of the constraint on the energy. 

We applied our method to four illustrative millisecond pulsar binary systems, for which we consistently modelled the companion atmosphere to provide a satisfactory fit to the available data.  We interpreted the presence or absence of day/night temperature modulation in the atmosphere in light of our simulated particle shower results. These calculations do not take into account the feedback of particle energy deposition on the atmosphere temperature. 

We find that companions with cool night sides showing strong temperature modulation provide the most interesting lower limits on the particle energy in the pulsar wind. For example, the evidence of deep heating in PSR B1957+20 constrains the particle energy to $\gtrsim 100\,$TeV at distances of $2.5\,R_\odot$ from the pulsar. In contrast, because of their high initial temperatures, redback systems are not favorable sites to constrain the composition of pulsar winds in this framework. One should caution that these results neglect stellar evolution and suppose that the system was in the currently observed configuration, but with a cool companion, before irradiation set in. 
This study of the irradiation of the cool-side atmosphere can be seen as a gedankenexperiment, which sets the grounds for further, more detailed studies.

On the other hand, a comparison to the day-side model shows that the high temperature difference once achieved can be maintained by a flux of particles with low energy of the order of $100\,$MeV. This latter result is general for all systems examined in this work, and is based on the state of the atmosphere as it is presently observed, which avoids uncertainties of stellar evolution.

Working on populations and specific binary systems with various orbital periods, it is thus possible to gain more information (number of impinging particles and energy distribution) on the content of the pulsar wind, at different distances from the pulsar.

We find that in some systems, the absence of modulation should result in shallow heating, which could be observed as inverted temperature and emission/absorption lines if detailed spectral measurements could be made. If such signatures are not observed, interpretations would require to invoke a system which is not tidally locked, or with poor irradiation efficiency due to large distances, or with heat deposited in an efficiently convective zone, deep in the companion atmosphere.

In future steps, dedicated modelings of specific companion atmospheres, including feedback from deep heating and system evolution would be needed to refine the estimates of the photospheric depths. Furthermore, more photometric and spectroscopic observations of millisecond binary systems with high resolutions are desired to increase the statistic and test the proposed idea to probe pulsar winds.

\section*{Acknowledgments}
First and foremost, we are very grateful to Sterl Phinney for initiating this project, and for his valuable comments on this manuscript. We would like to thank the anonymous referee for the detailed comments and suggestions, which helped improve the paper substantially. We acknowledge highly profitable discussions at the {\it Entretiens Pulsars} (supported by PNHE), in particular with David Smith, Lucas Guillemot, and Fabrice Mottez. We are also thankful to Jean Heyvaerts, J\'er\^ome P\'etri, Guillaume Dubus, Beno\^it Cerutti, Claire Gu\'epin and Lorenzo Cazon for fruitful discussions.  
This work is supported by the APACHE grant (ANR-16-CE31-0001) of the French Agence Nationale de la Recherche and the MINCYT (Argentina) through Grant No. PICT 2016-1128.

\bibliography{biblio}

\begin{thebibliography}{}
\makeatletter
\relax
\def\mn@urlcharsother{\let\do\@makeother \do\$\do\&\do\#\do\^\do\_\do\%\do\~}
\def\mn@doi{\begingroup\mn@urlcharsother \@ifnextchar [ {\mn@doi@}
  {\mn@doi@[]}}
\def\mn@doi@[#1]#2{\def\@tempa{#1}\ifx\@tempa\@empty \href
  {http://dx.doi.org/#2} {doi:#2}\else \href {http://dx.doi.org/#2} {#1}\fi
  \endgroup}
\def\mn@eprint#1#2{\mn@eprint@#1:#2::\@nil}
\def\mn@eprint@arXiv#1{\href {http://arxiv.org/abs/#1} {{\tt arXiv:#1}}}
\def\mn@eprint@dblp#1{\href {http://dblp.uni-trier.de/rec/bibtex/#1.xml}
  {dblp:#1}}
\def\mn@eprint@#1:#2:#3:#4\@nil{\def\@tempa {#1}\def\@tempb {#2}\def\@tempc
  {#3}\ifx \@tempc \@empty \let \@tempc \@tempb \let \@tempb \@tempa \fi \ifx
  \@tempb \@empty \def\@tempb {arXiv}\fi \@ifundefined
  {mn@eprint@\@tempb}{\@tempb:\@tempc}{\expandafter \expandafter \csname
  mn@eprint@\@tempb\endcsname \expandafter{\@tempc}}}

\bibitem[\protect\citeauthoryear{Abdo et~al.}{Abdo et~al.}{2009}]{Abdo_2009}
Abdo A.~A.,  et~al., 2009, \mn@doi [Astrophys. J. Suppl. S.]
  {10.1088/0067-0049/183/1/46}, 183, 46

\bibitem[\protect\citeauthoryear{{Abdo} et~al.,}{{Abdo}
  et~al.}{2013}]{2013ApJS..208...17A}
{Abdo} A.~A.,  et~al., 2013, \mn@doi [Astrophys. J. Suppl. S.]
  {10.1088/0067-0049/208/2/17}, \href
  {https://ui.adsabs.harvard.edu/abs/2013ApJS..208...17A} {208, 17}

\bibitem[\protect\citeauthoryear{{Acero} et~al.,}{{Acero}
  et~al.}{2015}]{2015ApJS..218...23A}
{Acero} F.,  et~al., 2015, \mn@doi [Astrophys. J. Suppl. S.]
  {10.1088/0067-0049/218/2/23}, \href
  {https://ui.adsabs.harvard.edu/abs/2015ApJS..218...23A} {218, 23}

\bibitem[\protect\citeauthoryear{Agostinelli et~al.,}{Agostinelli
  et~al.}{2003}]{GEANT}
Agostinelli S.,  et~al., 2003, \mn@doi [Nucl. Instrum. Methods Phys. Res. A]
  {10.1016/S0168-9002(03)01368-8}, 506, 250

\bibitem[\protect\citeauthoryear{{Archibald}, {Stairs}, {Ransom}
  et~al.}{{Archibald} et~al.}{2009}]{2009Sci...324.1411A}
{Archibald} A.~M.,  {Stairs} I.~H.,  {Ransom} S.~M.,   et~al., 2009, \mn@doi
  [Science] {10.1126/science.1172740}, \href
  {https://ui.adsabs.harvard.edu/abs/2009Sci...324.1411A} {324, 1411}

\bibitem[\protect\citeauthoryear{{Arons}}{{Arons}}{2003}]{Arons03}
{Arons} J.,  2003, \mn@doi [Astrophys. J.] {10.1086/374776}, \href
  {http://adsabs.harvard.edu/abs/2003ApJ...589..871A} {589, 871}

\bibitem[\protect\citeauthoryear{{Arons} \& {Tavani}}{{Arons} \&
  {Tavani}}{1993}]{Arons93}
{Arons} J.,  {Tavani} M.,  1993, \mn@doi [Astrophys. J.] {10.1086/172198},
  \href {http://adsabs.harvard.edu/abs/1993ApJ...403..249A} {403, 249}

\bibitem[\protect\citeauthoryear{Baglio, D'Avanzo, Campana, Zelati, Covino  \&
  Russell}{Baglio et~al.}{2016}]{Baglio:2016pow}
Baglio M.~C.,  D'Avanzo P.,  Campana S.,  Zelati F.~C.,  Covino S.,   Russell
  D.~M.,  2016, \mn@doi [Astron. Astrophys.] {10.1051/0004-6361/201628383},
  591, A101

\bibitem[\protect\citeauthoryear{{Bassa}, {van Kerkwijk}  \&
  {Kulkarni}}{{Bassa} et~al.}{2006}]{Bassa06}
{Bassa} C.~G.,  {van Kerkwijk} M.~H.,   {Kulkarni} S.~R.,  2006, \mn@doi
  [Astron. Astrophys.] {10.1051/0004-6361:20054316}, \href
  {http://adsabs.harvard.edu/abs/2006A%26A...450..295B} {450, 295}

\bibitem[\protect\citeauthoryear{{Bellm} et~al.,}{{Bellm}
  et~al.}{2013}]{2013AAS...22115410B}
{Bellm} E.,  et~al., 2013, in American Astronomical Society Meeting Abstracts
  \#221. p. 154.10, \url {http://adsabs.harvard.edu/abs/2013AAS...22115410B}

\bibitem[\protect\citeauthoryear{Bellm et~al.}{Bellm
  et~al.}{2016}]{Bellm:2015dfa}
Bellm E.~C.,  et~al., 2016, \mn@doi [Astrophys. J.]
  {10.3847/0004-637X/816/2/74}, 816, 74

\bibitem[\protect\citeauthoryear{{Bergeron}, {Saumon}  \&
  {Wesemael}}{{Bergeron} et~al.}{1995}]{Bergeron}
{Bergeron} P.,  {Saumon} D.,   {Wesemael} F.,  1995, \mn@doi [Astrophys. J.]
  {10.1086/175566}, \href
  {https://ui.adsabs.harvard.edu/abs/1995ApJ...443..764B} {443, 764}

\bibitem[\protect\citeauthoryear{Beronya, Karpova, Kirichenko, Zharikov,
  Zyuzin, Shibanov  \& Cabrera-Lavers}{Beronya et~al.}{2019}]{Beronya:2019lwc}
Beronya D.~M.,  Karpova A.~V.,  Kirichenko A.~{\relax Yu}.,  Zharikov S.~V.,
  Zyuzin D.~A.,  Shibanov {\relax Yu}.~A.,   Cabrera-Lavers A.,  2019, \mn@doi
  [Mon. Not. Roy. Astron. Soc.] {10.1093/MNRAS/stz607}, 485, 3715

\bibitem[\protect\citeauthoryear{Bhattacharyya et~al.}{Bhattacharyya
  et~al.}{2013}]{Bhattacharyya:2013ora}
Bhattacharyya B.,  et~al., 2013, \mn@doi [Astrophys. J.]
  {10.1088/2041-8205/773/1/L12}, 773, L12

\bibitem[\protect\citeauthoryear{Breton et~al.,}{Breton
  et~al.}{2013}]{Breton:2013ess}
Breton R.~P.,  et~al., 2013, \mn@doi [Astrophys. J.]
  {10.1088/0004-637X/769/2/108}, 769, 108

\bibitem[\protect\citeauthoryear{{Callanan}, {van Paradijs}  \&
  {Rengelink}}{{Callanan} et~al.}{1995}]{Callanan95}
{Callanan} P.~J.,  {van Paradijs} J.,   {Rengelink} R.,  1995, \mn@doi
  [Astrophys. J.] {10.1086/175229}, \href
  {http://adsabs.harvard.edu/abs/1995ApJ...439..928C} {439, 928}

\bibitem[\protect\citeauthoryear{Cho, Halpern  \& Bogdanov}{Cho
  et~al.}{2018}]{Cho2018}
Cho P.~B.,  Halpern J.~P.,   Bogdanov S.,  2018, \mn@doi [Astrophys. J.]
  {10.3847/1538-4357/aade92}, 866, 71

\bibitem[\protect\citeauthoryear{{Cholis}, {Hooper}  \& {Linden}}{{Cholis}
  et~al.}{2014}]{2014arXiv1407.5583C}
{Cholis} I.,  {Hooper} D.,   {Linden} T.,  2014, arXiv e-prints, \href
  {https://ui.adsabs.harvard.edu/abs/2014arXiv1407.5583C} {p. arXiv:1407.5583}

\bibitem[\protect\citeauthoryear{Crawford et~al.,}{Crawford
  et~al.}{2013}]{osti_22270898}
Crawford F.,  et~al., 2013, \mn@doi [Astrophys. J.]
  {10.1088/0004-637X/776/1/20}, 776

\bibitem[\protect\citeauthoryear{Draghis \& Romani}{Draghis \&
  Romani}{2018}]{Draghis:2018mkh}
Draghis P.,  Romani R.~W.,  2018, \mn@doi [Astrophys. J.]
  {10.3847/2041-8213/aad2db}, 862, L6

\bibitem[\protect\citeauthoryear{{Fang}, {Kotera}  \& {Olinto}}{{Fang}
  et~al.}{2012}]{Fang12}
{Fang} K.,  {Kotera} K.,   {Olinto} A.~V.,  2012, \mn@doi [Astrophys. J.]
  {10.1088/0004-637X/750/2/118}, \href
  {http://adsabs.harvard.edu/abs/2012ApJ...750..118F} {750, 118}

\bibitem[\protect\citeauthoryear{Fortin, Bejger, Haensel  \& Zdunik}{Fortin
  et~al.}{2016}]{Fortin:2014ufa}
Fortin M.,  Bejger M.,  Haensel P.,   Zdunik J.~L.,  2016, \mn@doi [Astron.
  Astrophys.] {10.1051/0004-6361/201424911}, 586, A109

\bibitem[\protect\citeauthoryear{{Fruchter}, {Stinebring}  \&
  {Taylor}}{{Fruchter} et~al.}{1988}]{1988Natur.333..237F}
{Fruchter} A.~S.,  {Stinebring} D.~R.,   {Taylor} J.~H.,  1988, \mn@doi [\nat]
  {10.1038/333237a0}, \href {http://adsabs.harvard.edu/abs/1988Natur.333..237F}
  {333, 237}

\bibitem[\protect\citeauthoryear{{Fruchter} et~al.,}{{Fruchter}
  et~al.}{1990}]{Fruchter90}
{Fruchter} A.~S.,  et~al., 1990, \mn@doi [Astrophys. J.] {10.1086/168502},
  \href {http://adsabs.harvard.edu/abs/1990ApJ...351..642F} {351, 642}

\bibitem[\protect\citeauthoryear{Gaisser, Engel  \& Resconi}{Gaisser
  et~al.}{2016}]{CRBook}
Gaisser T.,  Engel R.,   Resconi E.,  2016, Cosmic Rays and Particle Physics,
  2nd edn.
Cambrigde University Press, \mn@doi{10.1080/00107514.2017.1311376}

\bibitem[\protect\citeauthoryear{Gentile et~al.,}{Gentile
  et~al.}{2012}]{gentile_2012}
Gentile P.,  et~al., 2012, \mn@doi [Proceedings of the International
  Astronomical Union] {10.1017/S1743921312024234}, 8, 389

\bibitem[\protect\citeauthoryear{Haungs, Rebel  \& Roth}{Haungs
  et~al.}{2003}]{shower_rpp}
Haungs A.,  Rebel H.,   Roth M.,  2003, \mn@doi [Rep. Prog. Phys.]
  {10.1088/0034-4885/66/7/202}, 66, 1145

\bibitem[\protect\citeauthoryear{{Hoshino}, {Arons}, {Gallant}  \&
  {Langdon}}{{Hoshino} et~al.}{1992}]{Hoshino92}
{Hoshino} M.,  {Arons} J.,  {Gallant} Y.~A.,   {Langdon} A.~B.,  1992, \mn@doi
  [Astrophys. J.] {10.1086/171296}, \href
  {http://adsabs.harvard.edu/abs/1992ApJ...390..454H} {390, 454}

\bibitem[\protect\citeauthoryear{Huang, Kong, Takata, Hui, Lin  \& Cheng}{Huang
  et~al.}{2012}]{Huang_2012}
Huang R. H.~H.,  Kong A. K.~H.,  Takata J.,  Hui C.~Y.,  Lin L. C.~C.,   Cheng
  K.~S.,  2012, \mn@doi [Astrophys. J.] {10.1088/0004-637x/760/1/92}, 760, 92

\bibitem[\protect\citeauthoryear{{Hummer} \& {Mihalas}}{{Hummer} \&
  {Mihalas}}{1988}]{1988ApJ...331..794H}
{Hummer} D.~G.,  {Mihalas} D.,  1988, \mn@doi [Astrophys. J.] {10.1086/166600},
  \href {http://adsabs.harvard.edu/abs/1988ApJ...331..794H} {331, 794}

\bibitem[\protect\citeauthoryear{{Kao}, {Hallinan}, {Pineda}, {Stevenson}  \&
  {Burgasser}}{{Kao} et~al.}{2018}]{Kao18}
{Kao} M.~M.,  {Hallinan} G.,  {Pineda} J.~S.,  {Stevenson} D.,   {Burgasser}
  A.,  2018, \mn@doi [\apjs] {10.3847/1538-4365/aac2d5}, \href
  {https://ui.adsabs.harvard.edu/abs/2018ApJS..237...25K} {237, 25}

\bibitem[\protect\citeauthoryear{Kaplan, Bhalerao, van Kerkwijk, Koester,
  Kulkarni  \& Stovall}{Kaplan et~al.}{2013}]{Kaplan:2013hii}
Kaplan D.~L.,  Bhalerao V.~B.,  van Kerkwijk M.~H.,  Koester D.,  Kulkarni
  S.~R.,   Stovall K.,  2013, \mn@doi [Astrophys. J.]
  {10.1088/0004-637X/765/2/158}, 765, 158

\bibitem[\protect\citeauthoryear{Kaplan, Stovall, van Kerkwijk, Fremling  \&
  Istrate}{Kaplan et~al.}{2018}]{Kaplan:2018bwp}
Kaplan D.~L.,  Stovall K.,  van Kerkwijk M.~H.,  Fremling C.,   Istrate A.~G.,
  2018, \mn@doi [Astrophys. J.] {10.3847/1538-4357/aad54c}, 864, 15

\bibitem[\protect\citeauthoryear{Khechinashvili, Melikidze  \&
  Gil}{Khechinashvili et~al.}{2000a}]{Khechinashvili:2000as}
Khechinashvili D.,  Melikidze G.,   Gil J.,  2000a, \mn@doi [Astrophys. J.]
  {10.1086/309408}, 541, 335

\bibitem[\protect\citeauthoryear{{Khechinashvili}, {Melikidze}  \&
  {Gil}}{{Khechinashvili} et~al.}{2000b}]{Khechinashvili00}
{Khechinashvili} D.~G.,  {Melikidze} G.~I.,   {Gil} J.~A.,  2000b, \mn@doi
  [Astrophys. J.] {10.1086/309408}, \href
  {http://adsabs.harvard.edu/abs/2000ApJ...541..335K} {541, 335}

\bibitem[\protect\citeauthoryear{{Kirk}, {Lyubarsky}  \& {Petri}}{{Kirk}
  et~al.}{2009}]{Kirk09}
{Kirk} J.~G.,  {Lyubarsky} Y.,   {Petri} J.,  2009, in {Becker} W.,  ed.,
  Astrophysics and Space Science Library Vol. 357, Astrophysics and Space
  Science Library. p.~421 (\mn@eprint {} {arXiv:astro-ph/0703116}),
  \mn@doi{10.1007/978-3-540-76965-1$\_$16}

\bibitem[\protect\citeauthoryear{{Kluzniak}, {Ruderman}, {Shaham}  \&
  {Tavani}}{{Kluzniak} et~al.}{1988}]{Kluzniak88}
{Kluzniak} W.,  {Ruderman} M.,  {Shaham} J.,   {Tavani} M.,  1988, \mn@doi
  [Nature] {10.1038/334225a0}, \href
  {http://adsabs.harvard.edu/abs/1988Natur.334..225K} {334, 225}

\bibitem[\protect\citeauthoryear{{Kotera}, {Amato}  \& {Blasi}}{{Kotera}
  et~al.}{2015}]{Kotera15}
{Kotera} K.,  {Amato} E.,   {Blasi} P.,  2015, \mn@doi [J. Cosmol. Astropart.
  Phys.] {10.1088/1475-7516/2015/08/026}, \href
  {http://adsabs.harvard.edu/abs/2015JCAP...08..026K} {8, 026}

\bibitem[\protect\citeauthoryear{{Kotera}, {Mottez}, {Voisin}  \&
  {Heyvaerts}}{{Kotera} et~al.}{2016}]{Kotera2016}
{Kotera} K.,  {Mottez} F.,  {Voisin} G.,   {Heyvaerts} J.,  2016, \mn@doi
  [Astron. Astrophys.] {10.1051/0004-6361/201628116}, \href
  {http://adsabs.harvard.edu/abs/2016A%26A...592A..52K} {592, A52}

\bibitem[\protect\citeauthoryear{{Kowalski}}{{Kowalski}}{2014}]{Kowalski2014}
{Kowalski} P.~M.,  2014, \mn@doi [Astron. Astrophys.]
  {10.1051/0004-6361/201424242}, \href
  {https://ui.adsabs.harvard.edu/abs/2014A%26A...566L...8K} {566, L8}

\bibitem[\protect\citeauthoryear{Langer, Tauris  \& Kramer}{Langer
  et~al.}{2012}]{Tauris12}
Langer N.,  Tauris T.~M.,   Kramer M.,  2012, \mn@doi [Mon. Not. Roy. Astron.
  Soc.] {10.1111/j.1365-2966.2012.21446.x}, 425, 1601

\bibitem[\protect\citeauthoryear{{Lemoine}, {Kotera}  \& {P{\'e}tri}}{{Lemoine}
  et~al.}{2015}]{Lemoine15}
{Lemoine} M.,  {Kotera} K.,   {P{\'e}tri} J.,  2015, \mn@doi [J. Cosmol.
  Astropart. Phys.] {10.1088/1475-7516/2015/07/016}, \href
  {http://adsabs.harvard.edu/abs/2015JCAP...07..016L} {7, 016}

\bibitem[\protect\citeauthoryear{Li, Halpern  \& Thorstensen}{Li
  et~al.}{2014}]{Li:2014tka}
Li M.,  Halpern J.~P.,   Thorstensen J.~R.,  2014, \mn@doi [Astrophys. J.]
  {10.1088/0004-637X/795/2/115}, 795, 115

\bibitem[\protect\citeauthoryear{Li et~al.,}{Li et~al.}{2018}]{Li18}
Li K.-L.,  et~al., 2018, \mn@doi [Astrophys. J.] {10.3847/1538-4357/aad243},
  863, 194

\bibitem[\protect\citeauthoryear{Linares, Shahbaz  \& Casares}{Linares
  et~al.}{2018}]{Linares:2018ppq}
Linares M.,  Shahbaz T.,   Casares J.,  2018, \mn@doi [Astrophys. J.]
  {10.3847/1538-4357/aabde6}, 859, 54

\bibitem[\protect\citeauthoryear{Lyne et~al.,}{Lyne et~al.}{2013}]{Lyne2013}
Lyne A.,  et~al., 2013, \mn@doi [Mon. Not. Roy. Astron. Soc.]
  {10.1093/MNRAS/sts657}, 430, 571

\bibitem[\protect\citeauthoryear{{Manchester}, {Hobbs}, {Teoh}  \&
  {Hobbs}}{{Manchester} et~al.}{2005}]{Manchester05}
{Manchester} R.~N.,  {Hobbs} G.~B.,  {Teoh} A.,   {Hobbs} M.,  2005, VizieR
  Online Data Catalog, \href
  {http://adsabs.harvard.edu/abs/2005yCat.7245....0M} {7245, 0}

\bibitem[\protect\citeauthoryear{{Nice}, {Splaver}, {Stairs}, {L{\"o}hmer},
  {Jessner}, {Kramer}  \& {Cordes}}{{Nice} et~al.}{2005}]{2005ApJ...634.1242N}
{Nice} D.~J.,  {Splaver} E.~M.,  {Stairs} I.~H.,  {L{\"o}hmer} O.,  {Jessner}
  A.,  {Kramer} M.,   {Cordes} J.~M.,  2005, \mn@doi [Astrophys. J.]
  {10.1086/497109}, \href
  {https://ui.adsabs.harvard.edu/abs/2005ApJ...634.1242N} {634, 1242}

\bibitem[\protect\citeauthoryear{Nice, Stairs  \& Kasian}{Nice
  et~al.}{2008}]{Nice2008}
Nice D.~J.,  Stairs I.~H.,   Kasian L.~E.,  2008, \mn@doi [AIP Conference
  Proceedings] {10.1063/1.2900273}, 983, 453

\bibitem[\protect\citeauthoryear{{Orosz} \& {Hauschildt}}{{Orosz} \&
  {Hauschildt}}{2000}]{2000A&A...364..265O}
{Orosz} J.~A.,  {Hauschildt} P.~H.,  2000, Astron. Astrophys., \href
  {https://ui.adsabs.harvard.edu/abs/2000A%26A...364..265O} {364, 265}

\bibitem[\protect\citeauthoryear{{Patruno}}{{Patruno}}{2019}]{ApatrunoURL}
{Patruno} A.,  2019, Millisecond Pulsar Catalog, \href
  {https://apatruno.wordpress.com/about/millisecond-pulsar-catalogue/}
  {https://apatruno.wordpress.com/about/millisecond-pulsar-catalogue/}

\bibitem[\protect\citeauthoryear{{Phinney}, {Evans}, {Blandford}  \&
  {Kulkarni}}{{Phinney} et~al.}{1988}]{Phinney88}
{Phinney} E.~S.,  {Evans} C.~R.,  {Blandford} R.~D.,   {Kulkarni} S.~R.,  1988,
  \mn@doi [Nature] {10.1038/333832a0}, \href
  {http://adsabs.harvard.edu/abs/1988Natur.333..832P} {333, 832}

\bibitem[\protect\citeauthoryear{{Pletsch} \& {Clark}}{{Pletsch} \&
  {Clark}}{2015}]{Pletsch2015}
{Pletsch} H.~J.,  {Clark} C.~J.,  2015, \mn@doi [Astrophys. J.]
  {10.1088/0004-637X/807/1/18}, \href
  {https://ui.adsabs.harvard.edu/abs/2015ApJ...807...18P} {807, 18}

\bibitem[\protect\citeauthoryear{{Reynolds}, {Callanan}, {Fruchter}, {Torres},
  {Beer}  \& {Gibbons}}{{Reynolds} et~al.}{2007}]{Reynolds07}
{Reynolds} M.~T.,  {Callanan} P.~J.,  {Fruchter} A.~S.,  {Torres} M.~A.~P.,
  {Beer} M.~E.,   {Gibbons} R.~A.,  2007, \mn@doi [Mon. Not. Roy. Astron. Soc.]
  {10.1111/j.1365-2966.2007.11991.x}, \href
  {http://adsabs.harvard.edu/abs/2007Mon. Not. Roy. Astron. Soc..379.1117R}
  {379, 1117}

\bibitem[\protect\citeauthoryear{Rivera~Sandoval et~al.,}{Rivera~Sandoval
  et~al.}{2018}]{Sandoval:2017bog}
Rivera~Sandoval L.~E.,  et~al., 2018, \mn@doi [Mon. Not. Roy. Astron. Soc.]
  {10.1093/MNRAS/sty291}, 476, 1086

\bibitem[\protect\citeauthoryear{{Roberts}}{{Roberts}}{2013}]{Roberts2013}
{Roberts} M.~S.~E.,  2013, in {van Leeuwen} J.,  ed.,  IAU Symposium Vol. 291,
  Neutron Stars and Pulsars: Challenges and Opportunities after 80 years. pp
  127--132 (\mn@eprint {arXiv} {1210.6903}), \mn@doi{10.1017/S174392131202337X}

\bibitem[\protect\citeauthoryear{Rohrmann}{Rohrmann}{2001}]{Rohrmann:2001ku}
Rohrmann R.~D.,  2001, \mn@doi [Mon. Not. Roy. Astron. Soc.]
  {10.1046/j.1365-8711.2001.04298.x}, 323, 699

\bibitem[\protect\citeauthoryear{{Rohrmann}, {Serenelli}, {Althaus}  \&
  {Benvenuto}}{{Rohrmann} et~al.}{2002}]{2002MNRAS.335..499R}
{Rohrmann} R.~D.,  {Serenelli} A.~M.,  {Althaus} L.~G.,   {Benvenuto} O.~G.,
  2002, \mn@doi [Mon. Not. Roy. Astron. Soc.]
  {10.1046/j.1365-8711.2002.05644.x}, \href
  {https://ui.adsabs.harvard.edu/abs/2002MNRAS.335..499R} {335, 499}

\bibitem[\protect\citeauthoryear{Rohrmann, Althaus  \& Kepler}{Rohrmann
  et~al.}{2011}]{Rohrmann_MNRAS}
Rohrmann R.~D.,  Althaus L.~G.,   Kepler S.~O.,  2011, \mn@doi [Mon. Not. Roy.
  Astron. Soc.] {10.1111/j.1365-2966.2010.17716.x}, 411, 781

\bibitem[\protect\citeauthoryear{{Romani} \& {Sanchez}}{{Romani} \&
  {Sanchez}}{2016}]{Romani16}
{Romani} R.~W.,  {Sanchez} N.,  2016, \mn@doi [Astrophys. J.]
  {10.3847/0004-637X/828/1/7}, \href
  {https://ui.adsabs.harvard.edu/abs/2016ApJ...828....7R} {828, 7}

\bibitem[\protect\citeauthoryear{Romani \& Shaw}{Romani \&
  Shaw}{2011}]{Romani:2011xf}
Romani R.~W.,  Shaw M.~S.,  2011, \mn@doi [Astrophys. J.]
  {10.1088/2041-8205/743/2/L26}, 743, L26

\bibitem[\protect\citeauthoryear{{Romani}, {Filippenko}  \& {Cenko}}{{Romani}
  et~al.}{2015}]{Romani15}
{Romani} R.~W.,  {Filippenko} A.~V.,   {Cenko} S.~B.,  2015, \mn@doi
  [Astrophys. J.] {10.1088/0004-637X/804/2/115}, \href
  {https://ui.adsabs.harvard.edu/abs/2015ApJ...804..115R} {804, 115}

\bibitem[\protect\citeauthoryear{{Romani}, {Graham}, {Filippenko}  \&
  {Zheng}}{{Romani} et~al.}{2016}]{2016ApJ...833..138R}
{Romani} R.~W.,  {Graham} M.~L.,  {Filippenko} A.~V.,   {Zheng} W.,  2016,
  \mn@doi [Astrophys. J.] {10.3847/1538-4357/833/2/138}, \href
  {http://adsabs.harvard.edu/abs/2016ApJ...833..138R} {833, 138}

\bibitem[\protect\citeauthoryear{{Schroeder} \& {Halpern}}{{Schroeder} \&
  {Halpern}}{2014}]{2014ApJ...793...78S}
{Schroeder} J.,  {Halpern} J.,  2014, \mn@doi [Astrophys. J.]
  {10.1088/0004-637X/793/2/78}, \href
  {https://ui.adsabs.harvard.edu/abs/2014ApJ...793...78S} {793, 78}

\bibitem[\protect\citeauthoryear{{Shapiro} \& {Teukolsky}}{{Shapiro} \&
  {Teukolsky}}{1983}]{Shapiro83}
{Shapiro} S.~L.,  {Teukolsky} S.~A.,  1983, {Black holes, white dwarfs, and
  neutron stars: The physics of compact objects}.
Wiley, New York

\bibitem[\protect\citeauthoryear{{Stappers}, {Bessell}  \& {Bailes}}{{Stappers}
  et~al.}{1996}]{Stappers96}
{Stappers} B.~W.,  {Bessell} M.~S.,   {Bailes} M.,  1996, \mn@doi [Astrophys.
  J. Lett.] {10.1086/310397}, \href
  {http://adsabs.harvard.edu/abs/1996ApJ...473L.119S} {473, L119+}

\bibitem[\protect\citeauthoryear{{Stappers}, {van Kerkwijk}, {Lane}  \&
  {Kulkarni}}{{Stappers} et~al.}{1999}]{Stappers99}
{Stappers} B.~W.,  {van Kerkwijk} M.~H.,  {Lane} B.,   {Kulkarni} S.~R.,  1999,
  \mn@doi [Astrophys. J. Lett.] {10.1086/311795}, \href
  {http://adsabs.harvard.edu/abs/1999ApJ...510L..45S} {510, L45}

\bibitem[\protect\citeauthoryear{{Stappers}, {Gaensler}, {Kaspi}, {van der
  Klis}  \& {Lewin}}{{Stappers} et~al.}{2003}]{Stappers03}
{Stappers} B.~W.,  {Gaensler} B.~M.,  {Kaspi} V.~M.,  {van der Klis} M.,
  {Lewin} W.~H.~G.,  2003, \mn@doi [Science] {10.1126/science.1079841}, \href
  {http://adsabs.harvard.edu/abs/2003Sci...299.1372S} {299, 1372}

\bibitem[\protect\citeauthoryear{{Strader} et~al.,}{{Strader}
  et~al.}{2019}]{Strader18}
{Strader} J.,  et~al., 2019, \mn@doi [Astrophys. J.]
  {10.3847/1538-4357/aafbaa}, \href
  {https://ui.adsabs.harvard.edu/abs/2019ApJ...872...42S} {872, 42}

\bibitem[\protect\citeauthoryear{Tanabashi et~al.,}{Tanabashi
  et~al.}{2018}]{PhysRevD.98.030001}
Tanabashi M.,  et~al., 2018, \mn@doi [Phys. Rev. D]
  {10.1103/PhysRevD.98.030001}, 98, 030001

\bibitem[\protect\citeauthoryear{Tang et~al.}{Tang et~al.}{2014}]{Tang:2014nja}
Tang S.,  et~al., 2014, \mn@doi [Astrophys. J.] {10.1088/2041-8205/791/1/L5},
  791, L5

\bibitem[\protect\citeauthoryear{Thorstensen \& Armstrong}{Thorstensen \&
  Armstrong}{2005}]{Thorstensen_2005}
Thorstensen J.~R.,  Armstrong E.,  2005, \mn@doi [Astron. J.] {10.1086/431326},
  130, 759

\bibitem[\protect\citeauthoryear{{Verbiest}, {Weisberg}, {Chael}, {Lee}  \&
  {Lorimer}}{{Verbiest} et~al.}{2012}]{2012ApJ...755...39V}
{Verbiest} J.~P.~W.,  {Weisberg} J.~M.,  {Chael} A.~A.,  {Lee} K.~J.,
  {Lorimer} D.~R.,  2012, \mn@doi [Astrophys. J.] {10.1088/0004-637X/755/1/39},
  \href {https://ui.adsabs.harvard.edu/abs/2012ApJ...755...39V} {755, 39}

\bibitem[\protect\citeauthoryear{{van Kerkwijk}, {Bassa}, {Jacoby}  \&
  {Jonker}}{{van Kerkwijk} et~al.}{2005}]{vanKerkwijk:2004tm}
{van Kerkwijk} M.~H.,  {Bassa} C.~G.,  {Jacoby} B.~A.,   {Jonker} P.~G.,  2005,
  in {Rasio} F.~A.,  {Stairs} I.~H.,  eds,  Astronomical Society of the Pacific
  Conference Series Vol. 328, Binary Radio Pulsars. p.~357 (\mn@eprint {arXiv}
  {astro-ph/0405283})

\bibitem[\protect\citeauthoryear{{van Kerkwijk}, {Breton}  \& {Kulkarni}}{{van
  Kerkwijk} et~al.}{2011a}]{vanKerkwijk11}
{van Kerkwijk} M.~H.,  {Breton} R.~P.,   {Kulkarni} S.~R.,  2011a, \mn@doi
  [Astrophys. J.] {10.1088/0004-637X/728/2/95}, \href
  {http://adsabs.harvard.edu/abs/2011ApJ...728...95V} {728, 95}

\bibitem[\protect\citeauthoryear{van Kerkwijk, Breton  \& Kulkarni}{van
  Kerkwijk et~al.}{2011b}]{van_Kerkwijk_2011}
van Kerkwijk M.~H.,  Breton R.~P.,   Kulkarni S.~R.,  2011b, \mn@doi
  [Astrophys. J.] {10.1088/0004-637x/728/2/95}, 728, 95

\bibitem[\protect\citeauthoryear{van Staden \& Antoniadis}{van Staden \&
  Antoniadis}{2016}]{vanStaden:2016ubf}
van Staden A.~D.,  Antoniadis J.,  2016, \mn@doi [Astrophys. J.]
  {10.3847/2041-8213/833/1/L12}, 833, L12

\makeatother
\end{thebibliography}

\appendix

\section{Atmosphere modeling of PSR J0751+1807}\label{app:J0751}

\begin{figure}
\centering
\includegraphics[width=0.48\textwidth]{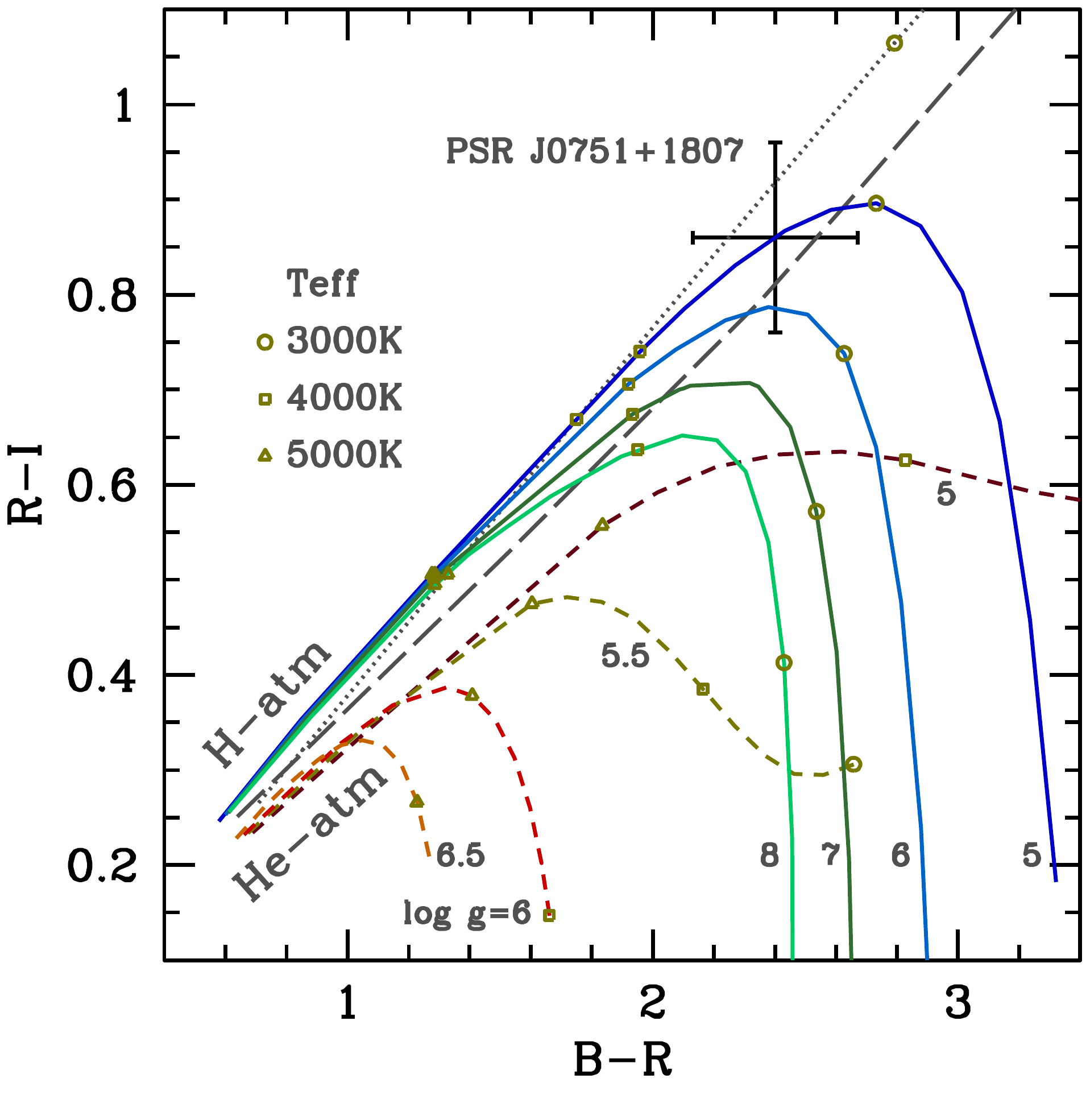}\\
\includegraphics[width=0.48\textwidth]{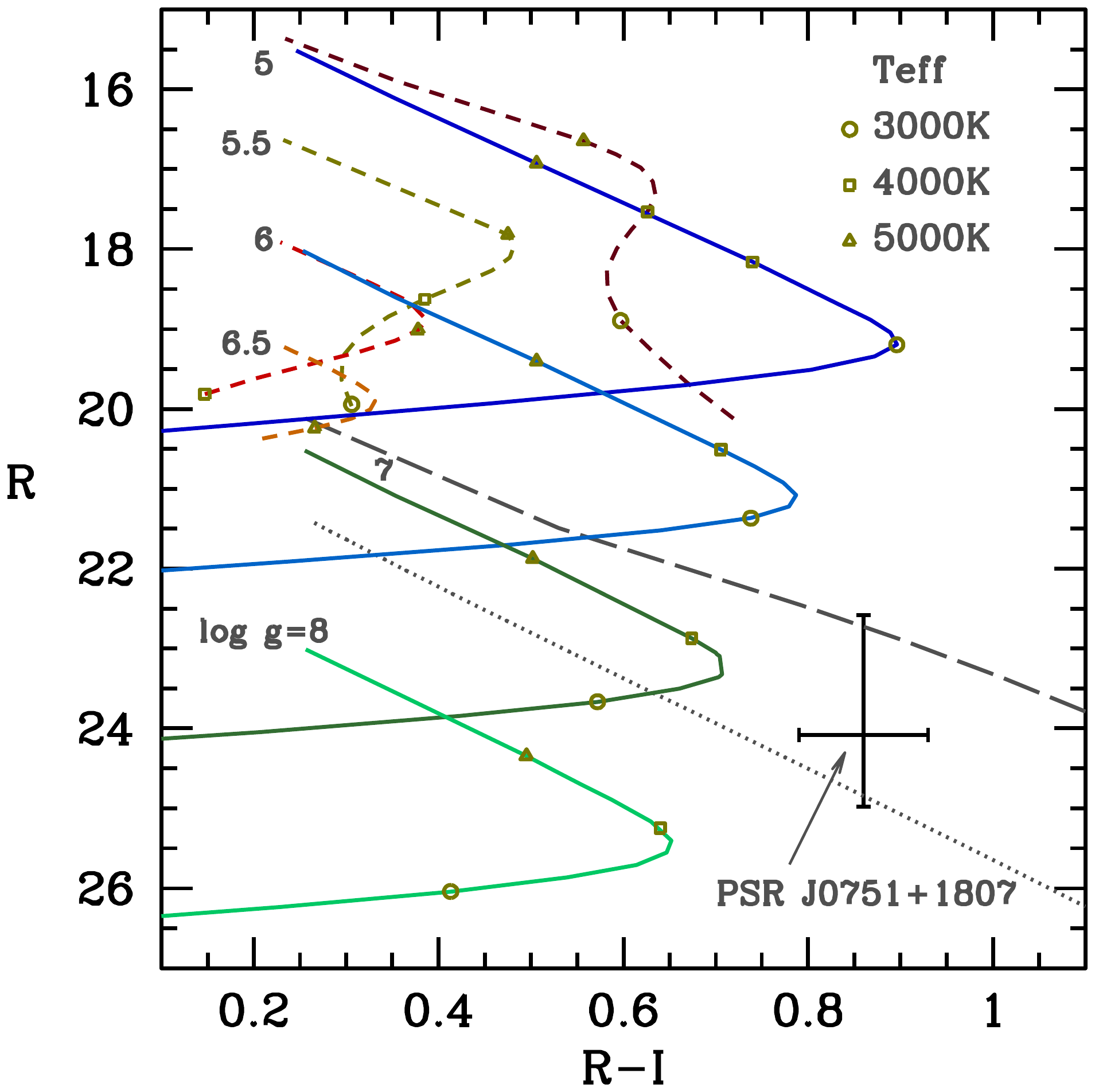}
\caption{Color-color (top) and color-magnitude (bottom) diagrams for the companion of PSR J0751+1807. Error bars indicate observations from Bassa et al. (2006). Lines denote theoretical curves from pure-H (solid), pure-He (dashed) models with updated CIA opacities, a blackbody model (dotted) and a He-pure model at $\log g=8$ (long dashed) without updated CIA opacities. Some values of $T_{\rm eff}$ and $\log g$ are indicated on the left plot. The magnitude $R$ increases with $\log g$ are indicated on the plots.}
\label{fig:J0751}
\end{figure}

The modeling of \cite{Bassa06} of the system PSR J0751+1807, a MSP with a white dwarf companion, is no longer valid, since new H and He opacities have been introduced since. In particular, collision-induced absorptions (CIA) of Lyman-alpha red wing (e.g., \citealp{Rohrmann_MNRAS}) and He-He-He interactions \citep{Kowalski2014}, which are present in hydrogen and helium (cool) atmospheres, respectively, have non-negligible effects on the atmospheres, as we show in Fig.~\ref{fig:J0751}.

\cite{Bassa06} found that pure-helium atmosphere models (from \citealp{Bergeron}, dashed line in Fig.~\ref{fig:J0751}) fit the photometry of PSR J0751+1807 at $T_{\rm eff} \approx 4000\,$K. Current helium models, which include He-He-He CIA opacity, do not adjust the observed magnitudes and colors. The helium CIA opacity (which yields the turn off in the $R-I$ color) has a cubic dependency on the gas density, so that its effects on the emergent stellar spectrum quickly increases as $T_{\rm eff}$ drops. This behaviour weakens at low $\log g$. On the other hand, H atmosphere models can fit colors of PSR J0751+1807 (with very low surface gravity) but not reproduce its magnitudes. Not too surprisingly, a simple blackbody model (dotted lines) may fit the observations with $\log g=6.754$ (using $M_{\rm C}=0.12\,M_\odot$ and $d=400$\,pc) and $T_{\rm eff}=3700\,$K. If the mass and distance values are reasonably well estimated, it would imply that current H and
 He atmosphere physics is not complete (e.g., missing opacities in cool temperature and high densities), or that the atmosphere of the companion of PSR J0751+1807 has a different composition. In this sense, it could not be discarded that material processed by prior nuclear burning (in particular CNO burning) could eventually be present in the outer envelope and atmosphere of the remnant, thus inflicting changes in the emergent spectrum.

\label{lastpage}

\end{document}